\documentclass[floats,floatfix,showpacs,amssymb,aps,prd,twocolumn,superscriptaddress,nofootinbib,nolongbibliography,reprint]{revtex4-2}

\usepackage{textcomp}
\usepackage{amssymb,amsmath,mathtools,needspace,enumitem,etoolbox,graphicx,physics,microtype,afterpage,gensymb,tabularx,mathtools,nicefrac,multirow,bm}
\usepackage[dvipsnames,table]{xcolor}
\definecolor{linkcolor}{rgb}{0.0,0.3,0.5}
\usepackage[unicode, colorlinks=true, linkcolor=linkcolor, citecolor=linkcolor, filecolor=linkcolor,urlcolor=linkcolor, pdfusetitle]{hyperref}
\usepackage[all]{hypcap}
\usepackage[T1]{fontenc}
\usepackage[english]{babel}
\usepackage[utf8]{inputenc}
\usepackage{orcidlink}
\usepackage[normalem]{ulem}

\newcommand{\ssim}{\mathchar"5218\relax\,}

\newcommand{\mmax}{m_{\rm max}}
\newcommand{\mmin}{m_{\rm min}}

\makeatletter
\newcommand*{\balancecolsandclearpage}{\close@column@grid \cleardoublepage \twocolumngrid}
\makeatother

\newcommand{\milan}{\affiliation{Dipartimento di Fisica ``G. Occhialini'', Universit\'a degli Studi di Milano-Bicocca, Piazza della Scienza 3, 20126 Milano, Italy}}
\newcommand{\infn}{\affiliation{INFN, Sezione di Milano-Bicocca, Piazza della Scienza 3, 20126 Milano, Italy}}
\newcommand{\geneva}{\affiliation{Departement de Physique Theorique and Gravitational Wave Science Center, Universit\'e de Geneve, 24 quai Ernest Ansermet, 1211 Geneve 4, Switzerland}}
\newcommand{\mrs}{\affiliation{Aix-Marseille Universit\'e, Universit\'e de Toulon, CNRS, CPT, Marseille, France}}

\begin{document}
\title{
Forecasting the population properties of merging black holes 
}

\author{Viola De Renzis\texorpdfstring{$\,$}{ }\orcidlink{0000-0001-7038-735X}}
\email{v.derenzis@campus.unimib.it}
\milan \infn

\author{Francesco Iacovelli\texorpdfstring{$\,$}{ }\orcidlink{0000-0002-4875-5862}}
\geneva

\author{Davide Gerosa\texorpdfstring{$\,$}{ }\orcidlink{0000-0002-0933-3579}}
\milan \infn

 \author{Michele Mancarella\texorpdfstring{$\,$}{ }\orcidlink{0000-0002-0675-508X}}
 \mrs 

\author{Costantino Pacilio\texorpdfstring{$\,$}{ }\orcidlink{0000-0002-8140-4992}}
\milan \infn


\date{\today}

\begin{abstract}

Third-generation gravitational-wave detectors will observe up to millions of merging binary black holes. With such a vast dataset, stacking events into population analyses will arguably be more important than analyzing single sources. %
We present the first application of population-level Fisher-matrix forecasts tailored to third-generation gravitational-wave interferometers \raisebox{-1pt}{\href{https://github.com/CosmoStatGW/gwfast/tree/master/gwfast/population}{\includegraphics[width=10pt]{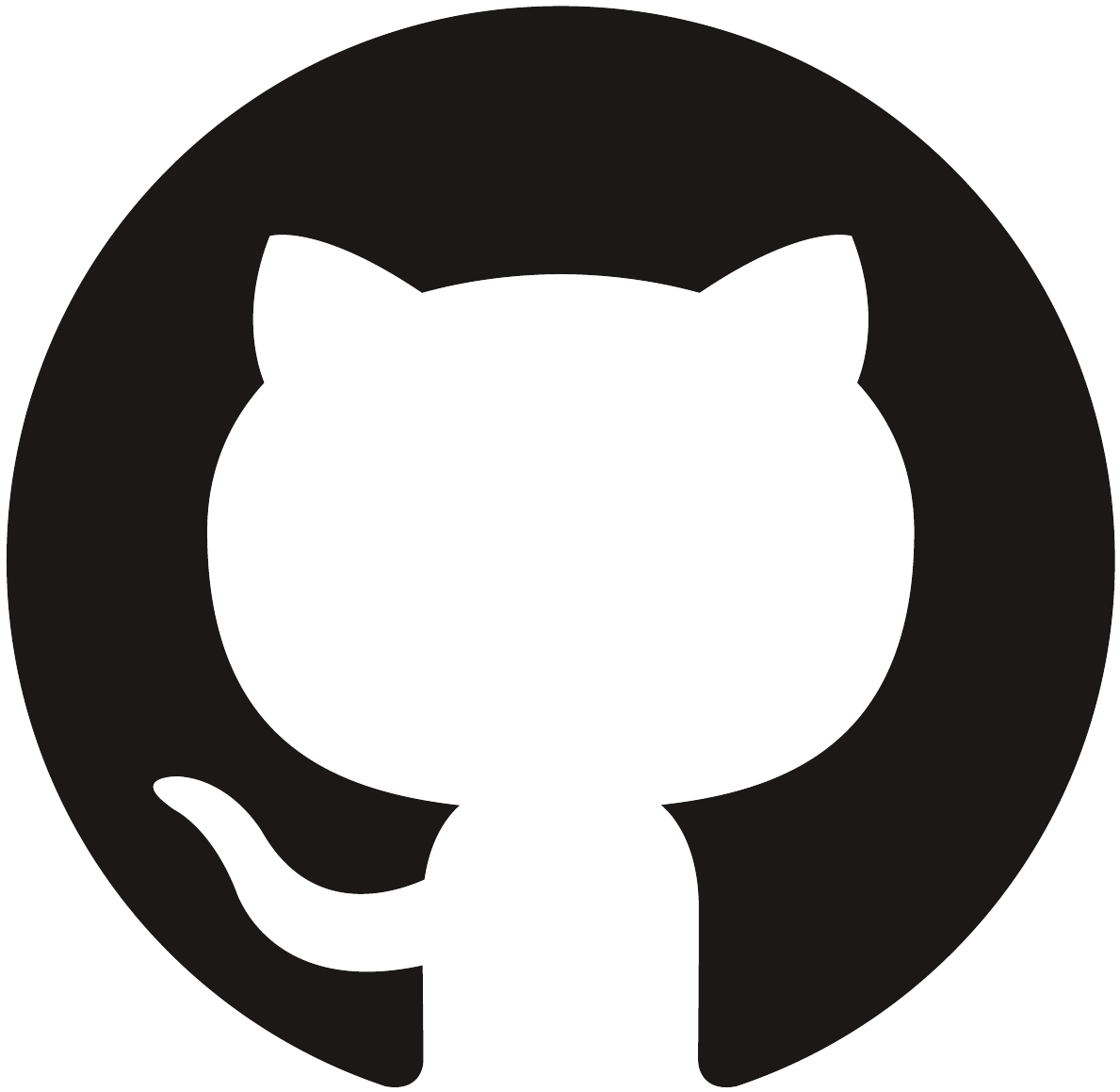}}}~\footnote{\url{https://github.com/CosmoStatGW/gwfast/tree/master/gwfast/population}}. We implement the formalism first derived by Gair {\it et al.} [\href{https://doi.org/10.1093/mnras/stac3560}{MNRAS 519, 2736 (2023)}] %
and explore how future experiments such as Einstein Telescope and Cosmic Explorer will constrain the distributions of black-hole masses, spins, and redshift. 
Third-generation detectors will be transformative, improving constraints on the population hyperparameters by several orders of magnitude compared to current data. 
At the same time, we highlight that a single third-generation observatory and a network of detectors  will deliver qualitatively similar performances.
Obtaining precise measurements of some population features (e.g. peaks in the mass spectrum)  will require only a few months of observations while others (e.g. the fraction of binaries with aligned spins) will instead require years if not decades. 
We argue population forecasts of this kind should be featured in white papers and feasibility studies aimed at developing the science case of future gravitational-wave interferometers.

\end{abstract}

\maketitle

\section{Introduction}

Gravitational-wave (GW) astronomy is quickly turning into a big-data discipline. We now have hundreds of detections, which will become thousands in a few years and millions in a few decades~\cite{2018LRR....21....3A,2023JCAP...07..068B}. The focus of GW inference is thus shifting from inferring the parameters of single events to those of the entire population of detected sources.

Current population analyses of LIGO/Virgo data~\cite{2019ApJ...882L..24A,2021ApJ...913L...7A,2023PhRvX..13a1048A} are providing precious information on the physical processes that drive black holes (BHs) and neutron stars to merger~\cite{2022PhR...955....1M,2021hgwa.bookE..16M,2021NatAs...5..749G}. In particular, there is strong indication that the merger rate of stellar-mass BHs 
decreases steeply for sources of masses $\gtrsim 60~M_\odot$ but piles up at $\ssim 35~M_\odot$. %
The spins of merging stellar-mass BHs show some preference for dimensionless magnitudes $\lesssim 0.3$ and some degree of co-alignment with the orbital angular momentum of the respective binaries. %
Besides marginal distributions, the growing GW dataset is now allowing us to explore correlations between parameters~\cite{2021ApJ...922L...5C,Adamcewicz:2022hce,Adamcewicz:2023mov,2022ApJ...932L..19B,2024PhRvD.109j3006H,2024A&A...684A.204R,Pierra:2024fbl,2024arXiv240616844H}.

State-of-the-art population analyses are tackled using hierarchical Bayesian statistics~\cite{2019MNRAS.486.1086M,2022hgwa.bookE..45V}. In short, one assumes an underlying inhomogeneous Poisson process and marginalizes over the parameters of the single events (e.g. masses, spins, etc.) while retaining information on the ``hyperparameters'' that describe the entire population of sources (e.g. mass cutoffs, spectral indexes, etc.). %
Current state-of-the-art implementations rely on recycling single-event posterior distributions for the computation of marginal integrals, as well as on injection campaigns to estimate selection biases via reweighted Monte Carlo integration~\cite{Tiwari:2017ndi,2019RNAAS...3...66F}. This requires careful handling of both finite-sampling and selection effects~\cite{2023MNRAS.526.3495T}. While under control for the time being, the feasibility of these techniques with much larger catalogs remains an open question.

The future of ground-based GW astronomy is under active planning, with flagship projects such as Einstein Telescope (ET)~\cite{Punturo:2010zz}
and Cosmic Explorer (CE)~\cite{2019BAAS...51g..35R} that promise to observe virtually all the BH binaries merging in the visible Universe, with detection rates of $\mathcal{O}(10^5 / {\rm yr})$~\cite{2022ApJ...941..208I,2023JCAP...07..068B}.
At present, running full Bayesian schemes on such large catalogs is both too costly and premature, if feasible at all. Science forecasts largely rely on the Fisher information matrix formalism~\cite{2008PhRvD..77d2001V}, which approximates the single-event likelihood in the high signal-to-noise ratio (SNR) limit. %
Fisher codes tailored to third-generation (3G) detectors include 
\textsc{GWFast}~\cite{2022ApJS..263....2I}, \textsc{GWBench}~\cite{2021CQGra..38q5014B}, \textsc{GWFish}~\cite{2023A&C....4200671D}, %
\textsc{TiDoFM~\cite{Li:2021mbo}},
and the code used in Ref.~\cite{Pieroni:2022bbh}. 
All such codes provide forecasts for the parameter estimation of single events rather than of population parameters.

On the other hand, in the third-generation (3G) 
era the focus of GW science will definitely be on populations. With the exception of prominent outliers, it is easy to image how the analysis of a single event will not be relevant when there are millions 
of similar sources sitting next to it. Despite this quite obvious statement, %
 current forecasts presented in ``colored books'' to funding agencies for the identification of the ET and CE science cases still rely on single-event analyses~\cite{2017CQGra..34d4001A,2020JCAP...03..050M,2021arXiv210909882E,2023JCAP...07..068B}.
We believe this is a severe limitation.

In this paper, we present the first population-level Fisher matrix implementation tailored to 3G detectors. The relevant equations have been developed by~\citeauthor{2023MNRAS.519.2736G}~\cite{2023MNRAS.519.2736G}, who first wrote down the Fisher-like expansion of the population likelihood including selection effects. Their expressions have so far only been applied to toy models. Here we significantly broaden the scope of the analysis to state-of-the-art parameterized population models, presenting detailed %
predictions for 3G detectors. We argue forecasts of this kind should be featured in future design documents to appropriately assess the outstanding science we will be able to deliver with facilities of the caliber of ET and CE.

Section~\ref{sec:formalism} introduces the Fisher formalism for both individual-event and population inference. Section~\ref{sec:methods} provides details on the implementation of our code, alongside a description of the population models used throughout this study. Section~\ref{sec:Results} presents our population-inference results, with a particular focus on the detector network as well as the astrophysical distributions of masses, spins, and redshift. Finally, Sec.~\ref{sec:conclusions} summarizes our main findings and discusses their implications.

\section{Formalism}
\label{sec:formalism}

In the following, $d$ indicates the data of a single GW signal, $\{d\}$ indicates data of a set of signals, $\theta$ indicates the parameters of individual events, and $\lambda$ indicates the hyperparameters describing the populations.
 
\subsection{Single-event inference}

Under the assumption that the noise is stationary and Gaussian, the likelihood of the event parameters is given by~\cite{Maggiore:2007ulw}
\begin{equation}\label{eq:single_ev_likelihood}
p(d|\theta)\propto \exp\left[-\frac{1}{2}\big(d-h(\theta)  |  d-h(\theta)\big)\right]\,,
 \end{equation}
where $h(\theta)$ is the gravitational signal %
and $(\cdot | \cdot)$ indicates the noise-weighted inner product. 
For a given BH binary with true parameters $\bar\theta$, the elements of the Fisher information matrix are given by
\begin{equation}
\Gamma_{\theta, ij} \equiv -{\bigg\langle \frac{\partial^2 \ln p(d|\theta)}{\partial\theta_i\partial\theta_j}\bigg\rangle  }\bigg|_{\theta=\bar\theta} 
= \left( \frac{\partial h}{\partial\theta_i}\bigg| \frac{\partial h}{\partial\theta_j} \right)\bigg|_{\theta=\bar\theta}  \,, \label{eq:FIM}
\end{equation}
where the expectation value $\langle \cdot \rangle$ is taken over noise realizations. 
The covariance matrix is given by $\Gamma_{\theta}^{-1}$ such that the errors on each parameter $\theta_i$ are given by the diagonal elements $(\Gamma^{-1}_{\theta})_{ii}^{\nicefrac{1}{2}}$.
The Fisher matrix provides a Gaussian approximation to the likelihood that becomes increasingly accurate in the high-SNR limit. This is equivalent to approximating the posterior distribution $p(\theta | d)$ if one assumes that the prior over $\theta$ is flat across the support of the likelihood itself. %

\subsection{Population inference}
\label{sec:Population_inference}

The chosen population model is encoded in the probability distribution $p_{\rm pop}(\theta | \lambda)$ that predicts the likelihood of compact objects merging with parameters $\theta$ for a given choice of the population $\lambda$. 
The population likelihood marginalized over the event rate is given by~\cite{2019MNRAS.486.1086M,2022hgwa.bookE..45V}
\begin{align}
p\left(\{d\}  |   {\lambda}\right) \propto p_{\rm det}({\lambda})^{-N_{\rm det}}\prod_{k=1}^{N_{\rm det}} \int p(d_k  |   {\theta}) p_{\rm pop}(\theta  | \lambda) \, \dd \theta\,,
\label{eq:populationLikelihood}
\end{align}
where $N_{\rm det}$ is the number of detected events in the catalog. 
While state-of-the-art, this expression assumes that GW events are statistically independent. This is an excellent approximation for LIGO/Virgo but a poorer one for ET and CE where multiple signals might overlap in the same stretch of data~\cite{2022PhRvD.105j4016P,2023MNRAS.523.1699J}.
Performing GW population analyses with overlapping signals is an unsolved problem. 

The term $p_{\rm det}({\lambda})$ in Eq.~\eqref{eq:populationLikelihood} is the fraction of detected events in the population described by $\lambda$, i.e.
\begin{equation}
p_{\rm det}(\lambda) = \int p_{\rm det}(\theta) p_{\rm pop}(\theta  |   \lambda) \, \dd \theta \leq 1\,,
\label{eq:pdet_lambda}
\end{equation}
where
$p_{\rm det}(\theta)$ is the probability of observing a merging binary with parameters $\theta$.
For simplicity and as common practice in the field, we use the SNR as a detection
statistics and set 
\begin{equation}
p_{\rm det}(\theta) =\frac{1}{2}\left\{ 1+ \text{erf}\left[\frac{\rho(\theta) -\rho_{\rm th}}{\sqrt{2}}\right] \right\}\,,
\label{eq:erfc}
\end{equation}
where $\rho_{\rm th}$ is a fixed threshold 
and $\rho(\theta)=\sqrt{(h(\theta)| h(\theta))}$ %
 is the optimal SNR. 
Equation~(\ref{eq:erfc}) includes the variance of the SNR due to noise realizations but not the effect of multiple detectors~\cite{2024CQGra..41l5002G}. We neglect this subtlety in favor of an efficient numerical implementation and use Eq.~\eqref{eq:erfc} also for detector networks, adding the SNRs of each instrument in quadrature. In particular, we set $\rho_{\rm th} =12$. %

The Fisher-matrix approximation of the likelihood in Eq.~\eqref{eq:populationLikelihood} has been worked out in Ref.~\cite{2023MNRAS.519.2736G}. Assuming a true population described by $\bar\lambda$, the elements of the hyper-Fisher matrix are
\begin{equation}
\Gamma_{\lambda,ij}  \equiv
- \left\langle \frac{\partial^2 \ln p\left(\{d\}  |   {\lambda}\right) }{\partial \lambda^i \partial \lambda^j} \right\rangle  \bigg|_{\lambda=\bar\lambda}\,,
\label{eq:exp_value_likelihood}
\end{equation}
which we write as a sum of five leading contributions%
\footnote{ The symbol $\Gamma_\lambda$ here indicates the hyper-Fisher matrix and not its single-event contribution as in Ref.~\cite{2023MNRAS.519.2736G}. With this notation, the components of $\Gamma_\lambda^{-1}$ directly provide the hyperparameters errors and correlations for a detection catalog of $N_{\rm det}$
events. 
Compared to notation of Ref.~\cite{2023MNRAS.519.2736G}, %
this is equivalent to adding a factor of 
$N_{\rm det}$ to the definition of $\Gamma_\lambda$. We have also factored out a common term $p_{\rm det}^{-1}(\lambda)$ from  $\Gamma_{\mathrm{I-V}}$.
}
\begin{equation}
\Gamma_{\lambda,ij} =\frac{N_{\det}}{p_{\rm det}({\lambda})} \,( \Gamma_{\mathrm{I}, ij} +\Gamma_{\mathrm{II},ij} +\Gamma_{\mathrm{III},ij} +\Gamma_{\mathrm{IV},ij} +\Gamma_{\mathrm{V},ij})\,,
\label{eq:5terms}
\end{equation}
where 
\begin{flalign} 
&\Gamma_{\mathrm{I},ij}\!=-\! \int \! \frac{\partial^2 \!\ln [p_{\rm pop}(\bar\theta  |  \lambda)p_{\rm det}^{-1}(\lambda)]}{\partial\lambda_i \partial\lambda_j}\Bigg|_{\lambda=\bar\lambda}  \!\!\!\!p_{\rm det}(\bar\theta) p_{\rm pop}(\bar\theta  |  \bar\lambda) \dd \bar\theta \label{eq:termI}\,, \\
&\Gamma_{\mathrm{II},ij} \!= \frac{1}{2} \! \int \! \frac{\partial^2\! \ln {\rm det}(\Gamma_\theta\!+\!H)}{\partial\lambda_i \partial\lambda_j} \Bigg|_{\lambda=\bar\lambda} \!\!\!\! p_{\rm det}(\bar\theta) p_{\rm pop}(\bar\theta  |  \bar\lambda) \dd \bar\theta \label{eq:termII} \,,\\
& \Gamma_{\mathrm{III},ij} \!=- \frac{1}{2} \! \int \! \frac{\partial^2\!\left[(\Gamma_\theta\!+\!H)^{-1}_{kl}\right] }{\partial\lambda_i \partial\lambda_j} \Bigg|_{\lambda=\bar\lambda}  \!\!\!\!\Gamma_{\theta,kl} \, p_{\rm pop}(\bar\theta  |  \bar\lambda) \dd \bar\theta\label{eq:termIII}  \,,\\
& \Gamma_{\mathrm{IV},ij} \!=- \! \int \! \frac{\partial^2 \!\left[ P_k(\Gamma_\theta\!+\!H)^{-1}_{kl}\right]}{\partial\lambda_i \partial\lambda_j} \Bigg|_{\lambda=\bar\lambda} \!\!\!\!D_{l} \, p_{\rm pop}(\bar\theta  |  \bar\lambda) \dd \bar\theta \label{eq:termIV}\,,\\
&\Gamma_{\mathrm{V},ij} \!=- \frac{1}{2}\! \int \! \frac{\partial^2 \!\left[ P_k (\Gamma_\theta\!+\!H)^{-1}_{kl} P_l \right]}{\partial\lambda_i \partial\lambda_j}  \Bigg|_{\lambda=\bar\lambda} \!\!\!\! p_{\rm det}(\bar\theta) p_{\rm pop}(\bar\theta  |  \bar\lambda) \dd \bar\theta\label{eq:termV} \,.
\end{flalign}
Here, $\Gamma_\theta$ is the single-event Fisher matrix defined in Eq.~\eqref{eq:FIM} and
\begin{align}
P_i&=\dfrac{\partial \ln p_{\rm pop}(\theta|\lambda)}{\partial \theta_i}\bigg|_{\theta=\bar\theta} \,,
\\
H_{ij}&= -\dfrac{\partial^2 \ln p_{\rm pop}(\theta|\lambda)}{\partial \theta_i \partial \theta_j}\bigg|_{\theta=\bar\theta} \,,
\\
D_{i}&= \dfrac{\partial p_{\rm det}(\theta)}{\partial\theta_i}\bigg|_{\theta=\bar\theta}= \dfrac{\partial p_{\rm det}}{\partial\rho}\dfrac{\partial\rho}{\partial\theta_i}\bigg|_{\theta=\bar\theta}\label{eq:derPDET}\,.
\end{align}
From these expressions, one can compute the covariance matrix $\Gamma_{\lambda}^{-1}$ and thus the errors on the population hyperparameters $(\Gamma^{-1}_{\lambda})_{ii}^{\nicefrac{1}{2}}$. 

In the limit where the uncertainties on the parameters of the individual sources are negligible, one has $\Gamma_{\theta}\to \infty$ which implies $\Gamma_{\mathrm{II-V}}\to 0$ because $\Gamma_{\theta}+H \sim \Gamma_{\theta}$, hence $(\Gamma_{\theta}+H)^{-1} \to 0$, and $\partial \Gamma_{\theta} / \partial \lambda = 0$.
The population Fisher matrix $\Gamma_\lambda$ can thus be approximated with the first term $\Gamma_{\mathrm{I}}$ of Eq.~\eqref{eq:termI}, which we expect to be the dominant contribution in our forecasting exercise (cf.~Ref.~\cite{2023MNRAS.519.2736G}).
Note that Eqs.~\eqref{eq:termI}-\eqref{eq:termV} are derived by averaging the population likelihood over noise realizations. This procedure ensures that the average impact of the noise bias is naturally included in the final Fisher expression, particularly in the term $\Gamma_{\mathrm{IV},ij}$ through its dependence on Eq.~\eqref{eq:derPDET}. For further details, we refer the reader to Ref~\cite{2023MNRAS.519.2736G}.

\subsection{Interpretation}

Equations~(\ref{eq:5terms}--\ref{eq:termV}) were computed assuming a Fisher-matrix description for the single-event likelihoods as well as large number of events, namely \mbox{$N_{\rm det}\to \infty$}~\cite{2023MNRAS.519.2736G}. The appropriate limit for a population Fisher analysis is that of many and well-measured GW signals. 
In this limit, the population Fisher matrix scales with the size of the catalog as $\Gamma_\lambda \propto N_{\rm det}$, 
cf.~Eqs.~\eqref{eq:pdet_lambda} and \eqref{eq:5terms} for any given population with efficiency $p_{\rm det}(\bar \lambda)$.

Crucially, the expressions above provide the population Fisher matrix only up to additional corrections scaling with inverse powers of $N_{\rm det}$. This implies that $\Gamma_\lambda$ is not guaranteed to be positive definite, thus casting some issues on the interpretation of its inverse as a covariance matrix. In practice, however, all diagonal terms $(\Gamma^{-1}_{\lambda})_{ii}$ are positive and those cases where the Fisher matrix $\Gamma_\lambda$ is non-positive definite are due to a small but negative eigenvalue. %
 If off-diagonal terms are needed (e.g. when drawing corner plots), we regularize $\Gamma_\lambda^{-1}$ by replacing the negative eigenvalue with its absolute value and projecting it back to the original coordinate space.

\section{Implementation}
\label{sec:methods}

\subsection{Approximants and detectors}

We implement the population Fisher-matrix formalism described above in the \textsc{Python} programming language, leveraging the infrastructure of the single-event Fisher code \textsc{GWfast}~\cite{2022ApJS..263....2I}.
All derivatives with respect to $\lambda$ are performed with automatic differentiation via \textsc{jax}~\cite{jax2018github}.

We use the {\sc IMRPhenomXPHM}~\cite{2020PhRvD.102f4001P,2020PhRvD.102f4002G,2021PhRvD.103j4056P} waveform model, 
and perform the single-event Fisher analysis in terms of the parameters
$\Tilde\theta=\{\mathcal{M}_z, \eta, \chi_1, \chi_2, \phi_{JL}, \phi_{12}, \theta_1, \theta_2, d_L, \alpha, \delta, \theta_{\rm JN}, \psi, t_c, \phi_c\}$. These are the detector-frame chirp mass $\mathcal{M}_z$, the symmetric mass ratio $\eta$, the dimensionless spin magnitudes $\chi_{1,2}$, tilt angles $\theta_{1,2}$, the azimuthal spin angle $\phi_{1,2}$, the azimuthal angle $\phi_{JL}$ 
between the total and orbital angular momenta, the luminosity distance $d_{L}$, the sky position coordinates $\alpha,\delta$, the polar angle $\theta_{JN}$ between total angular momentum and the line of sight, the polarization angle $\psi$, the time of coalescence $t_{c}$ and the phase at coalescence {$\phi_{c}$} (see Ref.~\cite{2022ApJ...941..208I}). We assume a flat $\Lambda$CDM model with \textsc{Planck18} parameters~\cite{2020A&A...641A...6P}.

We add a regularization matrix to the Fisher $\Gamma_{\Tilde\theta}$~\cite{2008PhRvD..77d2001V}, specifically targeting the sector associated with the
spin magnitudes and angular spin parameters; this serves two purposes: avoiding singular matrices $\Gamma_{\Tilde\theta}$ 
and incorporating information on the physical range of some parameters. 
Our regularization matrix is diagonal, with elements chosen as the inverse variances of the Gaussian 
distributions we impose. Specifically, the diagonal elements corresponding to the spin magnitudes, spin orientations, and azimuthal angles are \( \{  \sigma^{-2}_{\chi_1}, \sigma^{-2}_{\chi_2}, \sigma^{-2}_{\phi_{JL}}, \sigma^{-2}_{\phi_{12}}, \sigma^{-2}_{\theta_1}, \sigma^{-2}_{\theta_2} \} = \{ 1, 1, (2\pi)^{-2}, (2\pi)^{-2}, \pi^{-2}, \pi^{-2} \} \). All other elements are set to zero. 

Note the set of parameters $\Tilde\theta$ is not equivalent to the set $\theta$ entering our population analysis (see below), the difference being in the mass and redshift parameters.
We rotate the Fisher matrix $\Gamma_{\Tilde\theta}$ computed by {\sc GWFast} to obtain $\Gamma_\theta$. %
 In particular, one has $\Gamma_\theta=J^{T}\Gamma_{\Tilde\theta}J$ where $J=d\Tilde\theta/d\theta$ is the Jacobian of the rotation between the $\Tilde\theta$ and the $\theta$ parameters. In practice, this means we are assuming the likelihood is Gaussian also in the $\theta$ parameters; this is motivated by the large SNRs delivered by 3G detections.

We compare results obtained assuming a single, triangular shaped ET interferometer against a detector network composed of ET and two CE detectors~\cite{2023JCAP...07..068B,2021arXiv210909882E}. Specifically, we use the 10~km
noise curve for ET from Ref.~\cite{2023JCAP...07..068B} assuming a nominal location in Sardinia, Italy, and CE noise curves\footnote{These are available at 
\href{https://apps.et-gw.eu/tds/?r=18213}{apps.et-gw.eu/tds/?r=18213}
and
\href{https://dcc.cosmicexplorer.org/CE-T2000017/public}{dcc.cosmicexplorer.org/CE-T2000017/public}.

}
 assuming two interferometers with arm lengths of 40~km and 20~km in the USA~\cite{2022ApJ...931...22S,2021arXiv210909882E,2021CQGra..38q5014B}.
We set a minimum frequency of 2~Hz for ET and 5~Hz for CE.
The effect of the Earth rotation on the signal is taken into account when computing single-event Fisher matrices with \textsc{GWfast}.

\subsection{Population models} 
\label{popmodels}
The results presented in this work rely on a standard parametric population model $p_{\rm pop}(\theta|\lambda)$ borrowed from current LIGO/Virgo/KAGRA analyses~\cite{2023PhRvX..13a1048A}.
We tackle population inference on source-frame masses, spin magnitudes, spin tilts, and redshifts, i.e. $\theta = \{m_1,m_2,\chi_1,\chi_2,\theta_1,\theta_2,z, \Omega \}$ 
We assume that $p_{\rm pop}(\theta|\lambda)$ is separable over the source parameters as follows
\begin{align}
p_{\rm pop}(\theta|\lambda)& = p(m_1,m_2 | \lambda_m)p(\chi_1 | \lambda_\chi) p(\chi_2 | \lambda_\chi) 
\notag \\
& \times p(\theta_1 | \lambda_\theta) p(\theta_2 | \lambda_\theta) p(z|\lambda_z) p(\Omega)\,,
\end{align}
where $\lambda = \{\lambda_m, \lambda_z, \lambda_\chi, \lambda_\theta\}$. 
The symbol $\Omega$ collectively denotes all other parameters entering the GW approximant. Omitting some parameters in population inference is actually equivalent to assuming an astrophysical model where those parameters are distributed as in the Bayesian priors $p(\Omega)$ used in the underlying single-event analyses~\cite{2022hgwa.bookE..45V}. However, note that while the parameters $\Omega$ factor out in $p_{\rm pop}$, they do enter the detectability function $p_{\rm det}$.

We briefly summarize the population models used in this work as follows, see Table~\ref{tab:hyperpartable}. Appendix~\ref{app:popmodels} provides a more detailed description of these models. %

\begin{itemize}
\item \textit{Mass distribution}. We use a slightly edited version of the popular {\sc Power--Law + Peak} model~\cite{2023PhRvX..13a1048A,2018ApJ...863L..41F,2018ApJ...856..173T} with 9 free hyperparameters $\lambda_m=\{\alpha_m, \beta_q, \lambda_{\rm peak}, \mu_m,\sigma_m,m_{\rm min}, m_{\rm max}, \sigma_{\rm l},\sigma_{\rm h}\}$. The distribution of the primary mass $m_1$ is described by a power-law with slope $-\alpha_m$ extending between a minimum mass $m_{\rm min}-\sigma_{\rm l}$ and a maximum mass $m_{\rm max}+\sigma_{\rm h}$.
A Gaussian feature is included to account for a possible concentration of events just close to the pair-instability supernova mass gap. This Gaussian peak is characterized by weight $\lambda_{\rm peak}$, median $\mu_m$, and standard deviation $\sigma_m$. 
The secondary mass $m_2$ is modeled by a power-law with spectral index $\beta_q$. 
The distributions of $m_1$ and 
$m_2$ are smoothed over a range of masses $\sigma_{\rm l}$ ($\sigma_{\rm h}$) at the low (high) end of the mass function. Our smoothing prescription differ by that used elsewhere in the literature. In particular, we opted for polynomial smoothing functions for compatibility with the adopted automatic-differentiation strategy and numerical efficiency; see App.~\ref{app:popmodels}. 

\item \textit{Redshift distribution}. 
We use the so-called {\sc Madau--Dickinson} profile~\cite{2014ARA&A..52..415M} for the merger rate. This is parameterized by $\lambda_z=\{\alpha_z,\beta_z,z_{\rm p}\}$, namely the redshift $z_{\rm p}$ at which the rate peaks and the power-law indices $\alpha_z$ and $\beta_z$ that govern the rise and fall of rate at low and high redshifts, respectively.

\item \textit{Spin distribution}. The spin distribution is modeled after LIGO/Virgo's {\sc Default} model where $\lambda_\chi =\{ \alpha_\chi, \beta_\chi\}$
and
$\lambda_\theta =\{ \zeta,\sigma_t\}$~\cite{2023PhRvX..13a1048A,2017PhRvD..96b3012T}. The component spin magnitudes are independently drawn from a Beta distribution with 
shape parameters $\alpha_\chi$ and $\beta_\chi$.
The distribution of the spin tilt cosines have a uniform component (describing dynamically formed binaries with isotropically oriented spins) and a truncated Gaussian component with mean 1, weight $\zeta$ and standard deviation $\sigma_t$ (modeling BH binaries formed in isolated environment with preferentially aligned spins). 

\end{itemize}

Unless stated otherwise, we adopt fiducial values for the mass and spin hyperparameters $\lambda_m,\lambda_\chi,\lambda_\theta$ using the medians of the marginalized distributions from GWTC-3~\cite{2023PhRvX..13a1048A} and fiducial values of $\lambda_z$ from Ref.~\cite{2014ARA&A..52..415M}. In particular, we set $\alpha_m=3.4, \beta_q=1.1, \mmin=9.1~M_\odot,\mmax=87~M_\odot, \lambda_{\rm peak}=0.039, \mu_{\rm m}=34~M_\odot, \sigma_m=3.6~M_\odot, \sigma_{\rm l}=4~M_\odot,\sigma_{\rm h}=20~M_\odot, \alpha_z=2.7, \beta_z=3, z_{\rm p}=2, 
\alpha_\chi=1.6,\beta_\chi=4.12, 
\zeta=0.66, \sigma_t=1.5$. 
Note that our choice for the minimum mass $m_{\rm min}=9.1~M_\odot$ is different with respect to the median value of $\ssim5~M_\odot$ reported in Ref.~\cite{2023PhRvX..13a1048A}. This is because of the adopted smoothing function (see App.~\ref{app:popmodels}). In our implementation one has that the population distribution function is zero for $m \leq m_{\rm min} - \sigma_{\rm l}$. As a result, our parameter $m_{\rm min}$ is shifted by $\sigma_{\rm l}$ (which has fiducial value $4~M_\odot$) compared to that of Ref.~\cite{2023PhRvX..13a1048A}.

\subsection{Monte Carlo integrations and regularization}

We approximate Eqs.~\eqref{eq:pdet_lambda} and (\refeq{eq:termI}--\refeq{eq:termV}) with Monte Carlo integration. In particular, we cast those expressions as 
\begin{equation}
I(\bar\lambda) = \int X(\bar\theta,\bar\lambda) p_{\rm draw}(\bar\theta) \, \dd \bar\theta \simeq  \frac{1}{N_{\rm draw}}\sum_i^{N_{\rm draw}}X(\bar\theta_i,\bar\lambda)\,,
\label{eq:stdintegral}
\end{equation}
where $\bar\theta_i\sim p_{\rm draw}(\bar\theta)$ and $X(\bar\theta,\bar\lambda)$ is the ratio between the target distribution and $p_{\rm draw}(\bar\theta)$.

The injection distribution $p_{\rm draw}(\bar\theta)$ is chosen to improve convergence. 
For instance, it is convenient to use shallower power-laws (with spectral indexes closer to 0) to prevent undersampling in regions of the parameter space where $p_{\rm pop}(\bar \theta| \bar\lambda)$ is low but relevant for a specific element of $\bar\lambda$ (notably, this is the case of $m_{\rm max}$). For the mass model, we use a draw population with $\alpha_m=\beta_q=0.5$ and $\sigma_{\rm h}=20~M_\odot$. %
For all other hyperparamters, it is sufficient to draw samples using the true values $\bar \lambda$.

The subdominant terms $\Gamma_{\mathrm{II-V}}$ depend on the single-event Fisher matrices $\Gamma_\theta$ evaluated at the draws $\bar\theta _i$. If some of these sources have low SNRs and/or are located at the edge of the parameter space, this might result in occasionally large numerical errors on $\Gamma_\theta$, which in turn make the computation of $\Gamma_\lambda$ numerically unstable. As a regularization strategy, we restrict the computation of the Monte Carlo sum in Eq.~\eqref{eq:stdintegral} to the inner 95\% quantile of the evaluations $X(\bar\theta_i,\bar\lambda)$. At least for the \textsc{GWFast} implementation and using $N_{\rm draw}= \mathcal{O}(10^5)$, this is sufficient to obtain stable results. %
For additional regularization purposes, we also set $p_{\rm pop}( \theta| \lambda) = 0$ whenever %
this is $<10^{-12}$. This helps avoiding numerical instabilities %
when taking derivatives with respect to the hyperparameters $\lambda$ and we verified it does not affect our results in any meaningful way.%

In the following, the number of significant digits used when presenting the results is determined by the variance of the adopted Monte Carlo estimators.

\section{Results}
\label{sec:Results}

 \begin{figure*}[t]
    \includegraphics[width=0.95\textwidth]{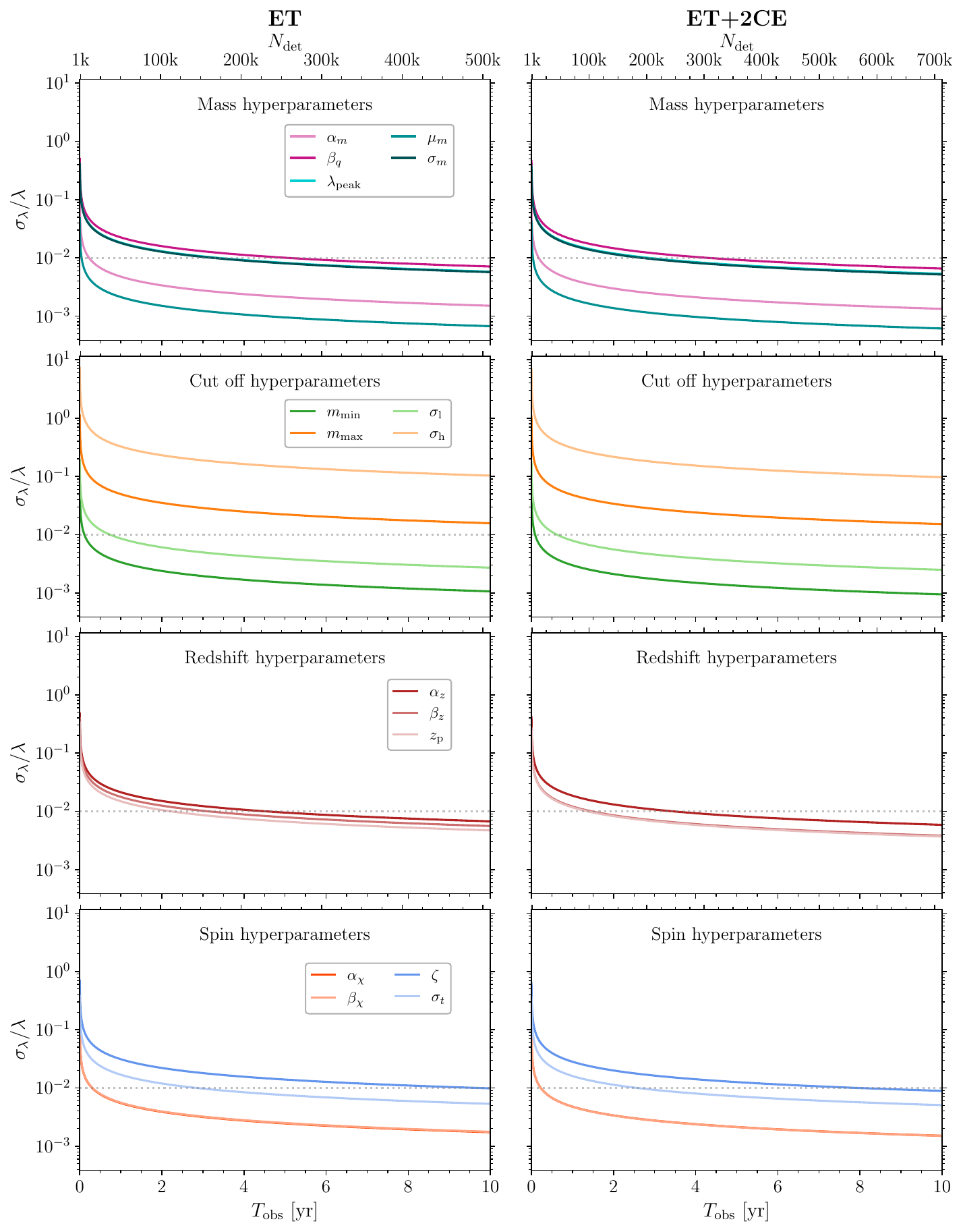}
    \caption{Relative errors of the model hyperparameters $\lambda$ as a function of the observation time (bottom $x$-axis) for ET (left panels) and ET+2CE (right panels) assuming our fiducial population model.
    The upper $x$-axis indicates the corresponding number of detected events, which depends on the chosen detector network. 
    The four rows correspond to the hyperparameters governing broad mass features, mass cutoffs, merger redshift evolution, and BH spins. To guide the eye, the dashed horizontal lines mark the threshold for percent-level accuracy, $\sigma_{\lambda}/\lambda = 1\%$. Some of these results are also reported in Table~ \ref{tab:ET_errors}.}   \label{fig:relative_errors}
\end{figure*}

\setlength{\tabcolsep}{4pt} %
\begin{table}
\centering 
\begin{tabular}{c||c|c||c|c} 
  & \multicolumn{2}{c||}{\textbf{ET}} & \multicolumn{2}{c}{\textbf{ET+2CE}} \\ \cline{2-5} 
  & $\sigma_\lambda/\lambda\;[10^{-3}]$ & $T_{1\%}$ [yr] & $\sigma_\lambda/\lambda\;[10^{-3}]$ & $T_{1\%}$ [yr]\\ \hline \hline 
$\alpha_{m}$ & $1.5$ & 0.2 & $1.3$ & 0.2 \\ 
$\beta_{q}$ & $7.1$ & 5.1 & $6.6$ & 4.3 \\ 
$\lambda_{\rm peak}$ & $5.8$ & 3.4 & $5.4$ & 2.9 \\ 
$\mu_{m}$ & $0.7$ & 0.1 & $0.6$ & 0.04 \\ 
$\sigma_{m}$ & $5.7$ & 3.3 & $5.2$ & 2.7 \\ \hline
$m_{\text{min}}$ & $1.1$ & 0.1 & $0.9$ & 0.1 \\ 
$m_{\text{max}}$ & $15.7$ & >10 & $15.1$ & >10 \\ 
$\sigma_{\rm l}$ & $2.7$ & 0.7 & $2.5$ & 0.6 \\ 
$\sigma_{\rm h}$ & $10.3$ & >10 & $9.7$ & >10 \\  \hline
$\alpha_{z}$ & $6.7$ & 4.5 & $5.9$ & 1.4 \\ 
$\beta_{z}$ & $5.6$ & 3.1 & $3.8$ & 0.2 \\ 
$z_{\rm p}$ & $4.7$ & 2.2 & $3.7$ & 1.4 \\ \hline
$\alpha_{\chi}$ & $1.7$ & 0.3 & $1.5$ & 0.2 \\ 
$\beta_{\chi}$ & $1.8$ & 0.3 & $1.5$ & 0.2 \\ 
$\zeta_{\chi}$ & $9.8$ & 9.7 & $8.9$ & 7.9 \\ 
$\sigma_{t}$ & $5.3$ & 2.8 & $5.1$ & 2.6
\end{tabular}
\caption{Summary results from our fiducial model, see also Fig.~\ref{fig:relative_errors}. For ET and ET+2CE, we report relative errors $\sigma_\lambda/\lambda$ after 10 years of observation as well as the required time to achieve a percent-level accuracy $\sigma_\lambda/\lambda = 1\%$. For readability, parameters are divided into four blocks (broad mass features, mass cutoffs, redshift, spins) as in Fig.~\ref{fig:relative_errors}. 
}
\label{tab:ET_errors} 
\end{table}

 \begin{figure*}[t]
    \includegraphics[width=\textwidth]{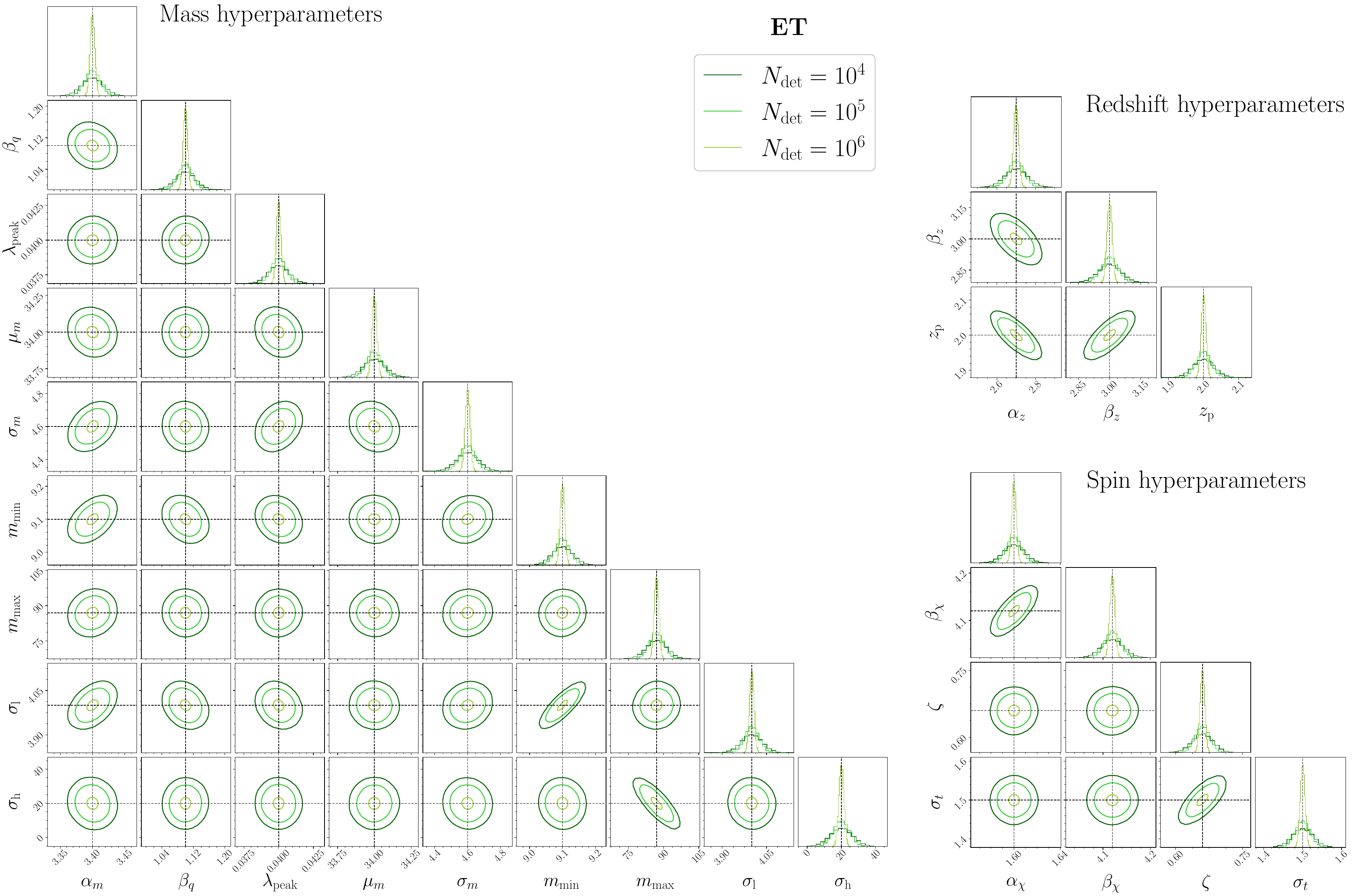}
    \caption{Hyperparameter uncertainties and their correlations assuming the ET detector and our fiducial population model. 
Contours indicate 90\% credible intervals and black dashed lines indicate the injected values.
We show results for catalogs of $N_{\rm det}= 10^4,10^5,10^6$ events (darker to lighter) corresponding to $T_{\rm obs}\simeq 0.2,2,19 $ years.
}   \label{fig:cornerET}
\end{figure*}

We now present our results, starting from a broad overview and then exploring forecasts for mass, redshift, and spins in more details.

 Figure~\ref{fig:relative_errors} shows the relative errors $\sigma_\lambda/\lambda$ of the 16 population hyperparameters as a function of the observation time for both the ET and ET+2CE configurations assuming our fiducial model. Results corresponding to $T_{\rm obs}=10$~yr are also reported in Table~\ref{tab:ET_errors}, together with time required to achieve 1\% accuracy in each of the hyperparameters. 
We divide our hyperparameters into four groups, separating those relative to broad mass features ($\alpha_m$, $\beta_q$, $\mu_m$, $\sigma_m$, $\lambda_{\rm peak}$), muss cutoffs ($m_{\rm max}$, $m_{\rm min}$, $\sigma_{\rm h}$, $\sigma_{\rm l}$), redshift ($z_{\rm p}$, $\alpha_z$, $\beta_z$) and spins ($\alpha_\chi$, $\beta_\chi$, $\zeta$, $\sigma_t$).
We consider a cumulative observation time $T_{\rm obs}\in[0,10]$~years and compute the corresponding number of detected events $N_{\rm det}$ %
as described in App.~\ref{app:popmodels}. %
For our fiducial model, an observation time of 10 years corresponds to $\ssim5\times10^5$ ($\ssim7\times10^5$) detected events and a detection efficiency $p_{\rm det}(\lambda)\simeq 69\%$ (96\%) for ET (ET+2CE). %

For the case of ET, Fig.~\ref{fig:cornerET} shows the joint distributions of the mass, spins, and redshift parameters, including their covariances.  A similar plot where we compare (with caveats) against current results from LIGO/Virgo is presented in App.~\ref{appLIGO}. 3G detectors will be transformative, improving population constraints by orders of magnitudes.

We also present some variations around the fiducial model, changing in particular the maximum mass $m_{\rm max}$, the peak of the redshift distribution $z_{\rm p}$, and the aligned/isotropic mixing fraction $\zeta$ for the spin directions. These results are presented in Figs.~\ref{fig:mmaxTobs}, \ref{fig:zp_figure}, \ref{fig:zetachi_figure}, and Table~\ref{tab:modelvariations}.

\subsection{Detector networks}
The most evident feature from Fig.~\ref{fig:relative_errors} is the scaling 
\begin{equation}
\frac{\sigma_\lambda}{\lambda}  \propto \sigma_\lambda \propto  \frac{1}{\sqrt{N}} \propto  \frac{1}{\sqrt{N_{\rm det}}} \propto \frac{1}{\sqrt{T_{\rm obs}}}\,,
\label{scaling}
\end{equation}
which is a direct consequence of Eqs.~\eqref{eq:pdet_lambda}, \eqref{eq:5terms}, and \eqref{eq:mergers2}.

From Fig.~\ref{fig:relative_errors} and Table~\ref{tab:ET_errors}, we predict that the ET and ET+2CE will measure the population properties of BH binaries equally well, at least qualitatively. While there are small quantitative differences, the overall picture is very similar between the two cases. Adding additional 3G detectors to the network will not be crucial for measuring the intrinsic population properties of merging BHs. 
This result is largely expected, for a few reasons: %
\begin{enumerate}[label=(\roman*)]
\item The overall scale of the problem is set by the number of detections. In the expressions above, this is indicated by the weight $p_{\rm det}(\bar\theta) p_{\rm pop}(\bar\theta  |  \bar\lambda)$ in the dominant $\Gamma_{\rm I}$ integral of Eq.~\eqref{eq:termI}. For a given intrinsic number of events $N$ (or equivalently $T_{\rm obs}$), upgrading from ET to ET+2CE improves population constraints by as little as 
\begin{equation}
1-\sqrt{\frac{ p_{\rm det, ET}(\bar \lambda) }{ p_{\rm det,ET+2CE}(\bar \lambda)}} \sim15\%\,.
\end{equation}
This is only an approximate estimate because the detectable population $p_{\rm det}(\bar\theta) p_{\rm pop}(\bar\theta  |  \bar\lambda)$ is integrated over an additional derivative term in Eq.~\eqref{eq:termI} and additional subdominant terms $\Gamma_{\rm II-V}$ are present.
\item While the individual-event measurement accuracies $\Gamma_\theta$ improve with multiple detectors, they enter the population Fisher matrix only through the subdominant terms $\Gamma_{\rm II-V}$. %
\item In this work and as currently done in most of the literature (but see Ref.~\cite{2020PhRvD.102j2004P} for a notable exception) we only consider the population of masses, spins, and redshift.
As is already the case for LIGO and Virgo, the accuracy of these parameters is largely dominated by the single most sensitive instrument in the network~\cite{2018LRR....21....3A}. On the other hand, it is well known that expanding the network greatly pays off in terms of source localization. %
\item We restrict our study to the \emph{global} population of sources and assume the BH merger rate roughly follows the star-formation rate, i.e. most of the events have $z\lesssim 2$. At these redshifts, the coverage provided by even a single %
ET instrument of triangular shape is essentially complete. This conclusion cannot be straightforwardly applied to putative sub-populations of sources at very high redshifts, e.g. those coming from 
population-3 stars~\cite{2014MNRAS.442.2963K} or primordial BHs~\cite{2018CQGra..35f3001S}.
 While the ET+2CE network has a larger detection horizon compared to ET alone, detecting binaries at cosmological distances does not imply an accurate inference of their source parameters~\cite{2022ApJ...931L..12N,2023PhRvD.107j1302M}. %

\end{enumerate}

\subsection{Mass distribution}

\begin{figure*}
    \includegraphics[width=0.75\textwidth]{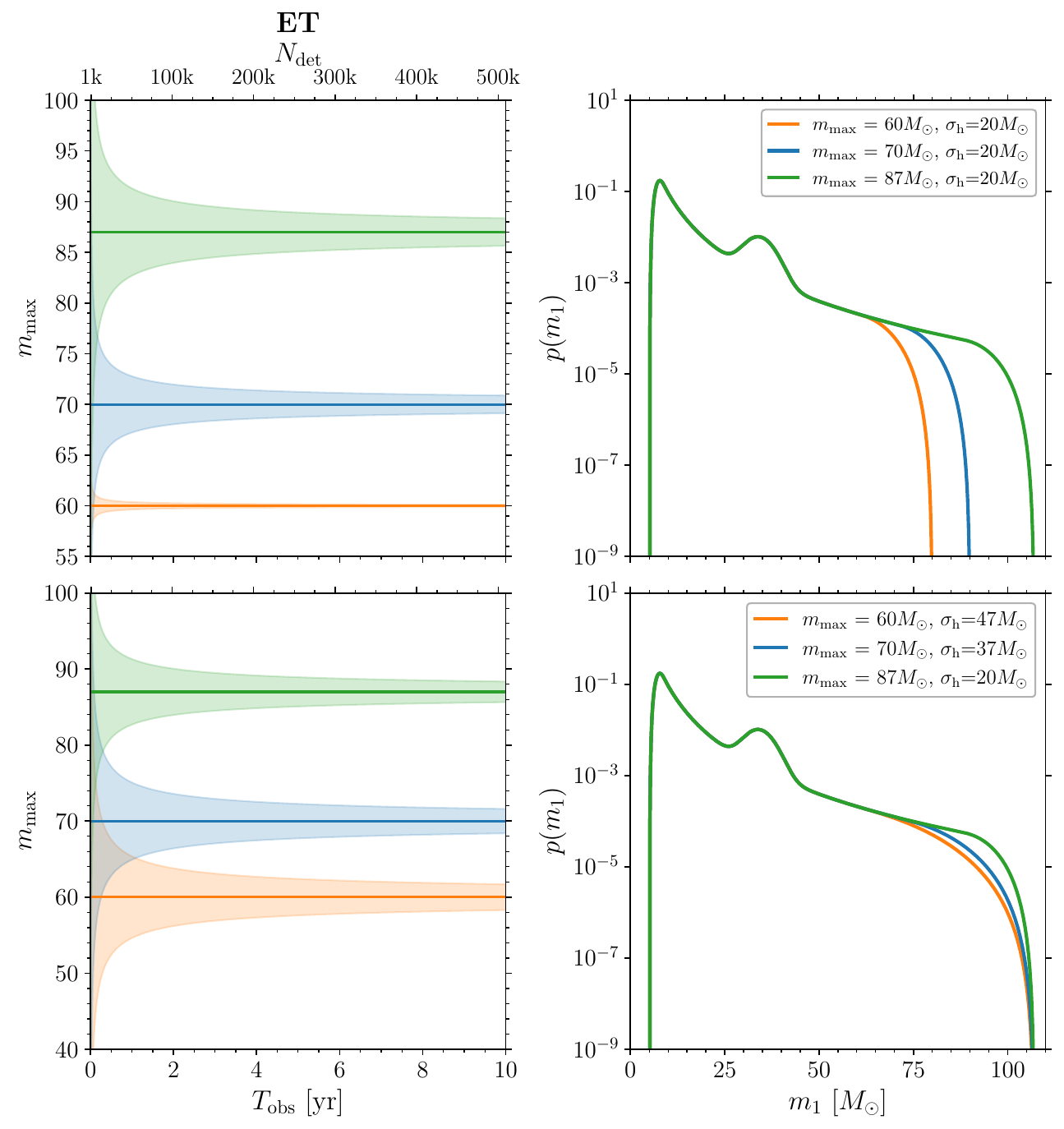}
    \caption{Measurability of upper edge of the mass spectrum as a function of the observation time $T_{\rm obs}$ (bottom $x$) axis for a single ET instrument. The corresponding number of detections is reported in the top $x$-axis. 
    Top (bottom) panels show model variations around the fiducial model where we vary the maximum mass $m_{\rm max}$ and fix (vary) the related smoothing parameter $\sigma_{\rm h}$. In both panels, the fiducial model is represented by the green curve with $m_{\rm max}=87~M_\odot$ and $\sigma_{\rm h}=20~M_\odot$. Left panels show the injected values (horizontal lines) and the related $1\sigma$ errors (shaded areas). For each model, the corresponding primary mass spectrum is illustrated in the right panels. %
       Related results are presented in Table~\ref{tab:modelvariations}.
}
            \label{fig:mmaxTobs}
\end{figure*}

The primary mass distribution of binary BHs inferred from GWTC-3 spans from a few solar masses up to around $90~M_\odot$. It shows a power-law behavior, with a Gaussian component at $\ssim35~M_\odot$, which is below the pair-instability supernova mass gap that is predicted by stellar evolution models for BHs with masses in the range $\ssim 40-120~M_\odot$~\cite{2016A&A...594A..97B,2019ApJ...882...36M,2019ApJ...882..121S,2017ApJ...836..244W}. The astrophysical origin of this excess is uncertain. %
The primary mass spectrum reveals another notable peak at $\ssim 10~M_\odot$, which has been suggested as an indicator of a significant contribution from isolated binary evolution~\cite{2018MNRAS.480.2011G,2020A&A...636A.104B,2022ApJ...940..184V}. %
The absence of a sharp cut-off for masses $\gtrsim40~M_\odot$ might point to evolutionary paths in dynamical environments, including the occurrence of hierarchical mergers and/or formation in environments like galactic nuclei~\cite{2016ApJ...831..187A,2020ApJ...898...25T,2021NatAs...5..749G}. %

Figure~\ref{fig:relative_errors}  and  Table~\ref{tab:ET_errors} show that hyperparameters such as the minimum mass $m_{\text{min}}$, the spectral index $\alpha_m$, and the position of the Gaussian peak $\mu_m$ in the primary mass spectrum are measured with high precision, reaching 1\% accuracy in just a few months of 3G operation. All of these are parameters that impact the population where the event rate is sufficiently large (i.e. close to the ``top'' of the underlying power-law structure). %
Measuring $m_{\text{min}}$ will test the occurrence of the putative mass gap between BHs and neutron stars, measuring  $\alpha_m$ will indicate whether the BH mass spectrum is  shallower or steeper than that of stars, and measuring $\mu_m$  will delineate the boundary between BHs formed through supernova explosion and those
potentially limited by pair instabilities,  
 shedding light on the underlying physics of core collapse. The association of the Gaussian peak with the pulsational pair-instability mechanism remains uncertain, see e.g. Refs.~\cite{2023MNRAS.526.4130H,2023MNRAS.520.5724B}.

The secondary-mass spectral index 
$\beta_q$ is measured less accurately than that of the primary mass $\alpha_m$. This is due to both $m_1$ being easier to constrain than $m_2$, but also to the specific functional form of $p_{\rm pop}(\theta|\lambda)$ used here, which is written as two independent contributions $p(m_1) p(m_2| m_1)$.  %

In contrast, parameters such as the maximum mass $m_{\text{max}}$ and the smoothing length $\sigma_{\rm h}$ are harder to measure because they affect the low-rate region of the population (at the ``bottom'' end of the underlying power-law). The accuracy of both these parameters does not go below 1\% accuracy even after 10 years of 3G observations.

The corner plot of Fig.~\ref{fig:cornerET} shows that, as expected, the cutoff parameters at both the lower and upper ends of the mass spectrum exhibit distinct correlations with their respective smoothing hyperparameters. Specifically, the minimum mass $m_{\rm min}$ shows a positive correlation with $\sigma_{\rm l}$, indicating that an increase in $m_{\rm min}$ can be compensated by a simultaneous increase in $\sigma_{\rm l}$ while still fitting the same population model. Conversely, the maximum mass $m_{\text{max}}$ displays a negative correlation with $\sigma_{\rm h}$; thus, an increase in $m_{\rm max}$ necessitates a corresponding decrease in $\sigma_{\rm h}$ to maintain the same probability density function $p_{\rm pop}(\theta | \lambda)$.

We also observe a positive correlation between $m_{\rm min}$ (and thus $\sigma_{\rm l}$) and the spectral index $\alpha_m$. This is because both a steeper power-law (larger $\alpha_m$) and a more extended mass spectrum (lower $m_{\rm min}$ ) predict a larger number of low-mass detections. %
We do not observe a similar correlation between $\alpha_m$ and $m_{\text{max}}$ because there are significantly fewer events at the upper end of the mass spectrum. %
A similar but weaker behavior is observed also for the secondary-mass
spectral index $\beta_q$.

The minimum mass $m_{\rm min}$ (and consequently $\sigma_{\rm l}$) is also correlated with the fraction of sources in the Gaussian component $\lambda_{\rm peak}$. %
As $m_{\rm min}$ and $\sigma_{\rm l}$ increase, the contribution provided by the power-law component decreases (the power-law is ``shorter'') and thus the weight of the Gaussian peak must increase.
We further observe a negative correlation between the spectral indices $\alpha_m$ and $\beta_q$ of the primary and secondary mass distribution, respectively. This can be understood as follows: a higher value for the primary mass spectral index, i.e. a steeper power-law, results in a larger number of light events, consequently flattening the secondary mass distribution, and thus lowering the value of $\beta_q$.

\begin{table}[t!]
\setlength{\tabcolsep}{3.7pt}
\begin{tabular}{l|c|c|c}

 & \multicolumn{3}{c}{$\sigma_\lambda~[10^{-2}]$}\\
\hline \hline
 & $m_{\rm max}=60~M_\odot$ & $m_{\rm max}=70~M_\odot$ & $\bm{m_{\rm max}=87~M_\odot}$ \\
 & $\sigma_{\rm h}=20~M_\odot$ &$\sigma_{\rm h}=20~M_\odot$ & $\bm{\sigma_{\rm h}=20~M_\odot}$ \\ \hline
$\alpha_{m}$ & 0.28 & 0.25 & 0.51 \\ 
$\beta_{q}$ & 0.71 & 0.80 & 0.78 \\ 
$\lambda_{\rm peak}$ & 0.02 & 0.02 & 0.02 \\ 
$\mu_{m}$ & 2.47 & 2.52 & 2.29 \\ 
$\sigma_{m}$ & 2.49 & 2.99 & 2.05 \\ 
$m_{\text{min}}$ & 1.02 & 0.95 & 0.97 \\ 
$m_{\text{max}}$ & 11.59 & 87.80 & 136.23 \\ 
$\sigma_{\rm l}$ & 1.15 & 1.08 & 1.08 \\ 
$\sigma_{\rm h}$ & 11.77 & 131.20 & 205.65 \\ \hline \hline
 & $m_{\rm max}=60~M_\odot$ & $m_{\rm max}=70~M_\odot$ & $\bm{m_{\rm max}=87~M_\odot}$ \\ 
 & $\sigma_{\rm h}=47~M_\odot$ &$\sigma_{\rm h}=37~M_\odot$ & $\bm{\sigma_{\rm h}=20~M_\odot}$ \\ \hline
$\alpha_{m}$ & 0.56 & 0.54 & 0.51 \\ 
$\beta_{q}$ & 0.82 & 0.81 & 0.78 \\ 
$\lambda_{\rm peak}$ & 0.02 & 0.02 & 0.02 \\ 
$\mu_{m}$ & 2.40 & 2.40 & 2.29 \\ 
$\sigma_{m}$ & 2.15 & 2.14 & 2.05 \\ 
$m_{\text{min}}$ & 0.94 & 0.93 & 0.97 \\ 
$m_{\text{max}}$ & 169.90 & 160.62 & 136.23 \\ 
$\sigma_{\rm l}$ & 1.03 & 1.03 & 1.08 \\ 
$\sigma_{\rm h}$ & 264.31 & 245.29 & 205.65 \\\hline \hline 
 & $z_{\rm p}=1$ &  $\bm{z_{\rm p}=2}$ & $z_{\rm p}=5$ \\ \hline  
$\alpha_{z}$ & 3.32 & 1.81 & 0.92 \\ 
$\beta_{z}$ & 1.35 & 1.67 & 2.95 \\ 
$z_{\rm p}$ & 0.70 & 0.94 & 1.78 \\ \hline \hline
 & $\zeta=0.1$ & $\bm{\zeta=0.66}$ & $\zeta=1$ \\ \hline
$\alpha_{\chi}$ & 0.27 & 0.28 & 0.28 \\ 
$\beta_{\chi}$ & 0.72 & 0.73 & 0.73 \\ 
$\zeta$ & 0.32 & 0.65 & 0.70 \\ 
$\sigma_{t}$ & 20.95 & 7.99 & 5.62
\end{tabular}
\caption{Summary results from a few model variations. We show absolute errors $\sigma_\lambda$ 
 after 10 years of observation with ET.
Starting from our fiducial model, we vary one or two parameters at a time. In particular, we explore variations of the maximum mass $m_{\rm max}$ (first block of cells), the maximum mass $m_{\rm max}$ and the smoothing paramenter $\sigma_{\rm h}$ (second block of cells), the peak of the redshift distribution $z_{\rm p}$ (third block of cells), and the fraction of binaries with aligned spins $\zeta$ (fourth block of cell). The fiducial model corresponds to $m_{\rm max}=87~M_\odot$, $z_{\rm p}=2$, and $\zeta=0.66$ as indicated in boldface. Related results are presented in Figs.~\ref{fig:mmaxTobs}, \ref{fig:zp_figure},~\ref{fig:zetachi_figure}. 
}
\label{tab:modelvariations}
\end{table}

In Fig.~\ref{fig:mmaxTobs} and Table~\ref{tab:modelvariations}, we present two model variations %
to study the measurability of the upper cutoff $m_{\rm max}$. The impact of these model variations on $p_{\rm det}(\lambda)$ is of $\mathcal{O}(10^{-4})$, such that a given value of $T_{\rm obs}$ corresponds to GW catalogs of essentially the same size $N_{\rm det}$.%

\begin{enumerate}[label=(\roman*)]

\item First, we vary the maximum mass $m_{\rm max}=60~M_\odot,\, 70~M_\odot,\, 87~M_\odot$ while keeping all other hyperparameters fixed to their fiducial values, including the high-mass smoothing $\sigma_{\rm h}$. %
We find that the uncertainties on the maximum mass increase with the value of the maximum mass itself. After $T_{\rm obs}=10$ years, ET will be able to constrain the maximum mass with absolute errors of $\pm[1.159,\, 8.780,\,  13.623]\times 10^{-1}~M_\odot$ for $m_{\rm max}=[60,\, 70,\, 87]~M_\odot$. Lowering the value of $m_{\rm max}$ shifts the cutoff closer to the region of the mass spectrum where the event rate is higher. Consequently, the greater density of observed events provides more statistical information, allowing for more precise population constraints. A lower value of the maximum mass also decreases the errors on the smoothing parameter $\sigma_{\rm h}$; this might be due to the correlation between $\sigma_{\rm h}$ and $m_{\rm max}$ highlighted above. %
 The errors on the other mass hyperparameters remain approximately constant under these model variations. 

\item Second, we simultaneously vary the maximum mass $m_{\rm max}=60~M_\odot ,\, 70~M_\odot,\, 87~M_\odot$ and the smoothing length $\sigma_{\rm h}=47~M_\odot,\, 37~M_\odot,\, 20~M_\odot$ such that their sum $m_{\rm max}+\sigma_{\rm h}=107~M_\odot$ remains constant. Contrary to the previous case, here we find that errors on the maximum mass and that of the smoothing parameter $\sigma_{\rm h}$ decrease with the value of $m_{\rm max}$. In this controlled experiment, lower values of $m_{\rm max}$ imply higher values of $\sigma_{\rm h}$ and thus a smoother high-mass end of the mass spectrum. Such prominent smoothing reduces the importance of the features near the cutoff, making it harder to pinpoint the exact location of $m_{\rm max}$. As a result, the uncertainty on both $m_{\rm max}$ and $\sigma_{\rm h}$ increases with decreasing~$m_{\rm max}$.

\end{enumerate}

\begin{figure*}
    \includegraphics[width=0.765\textwidth]{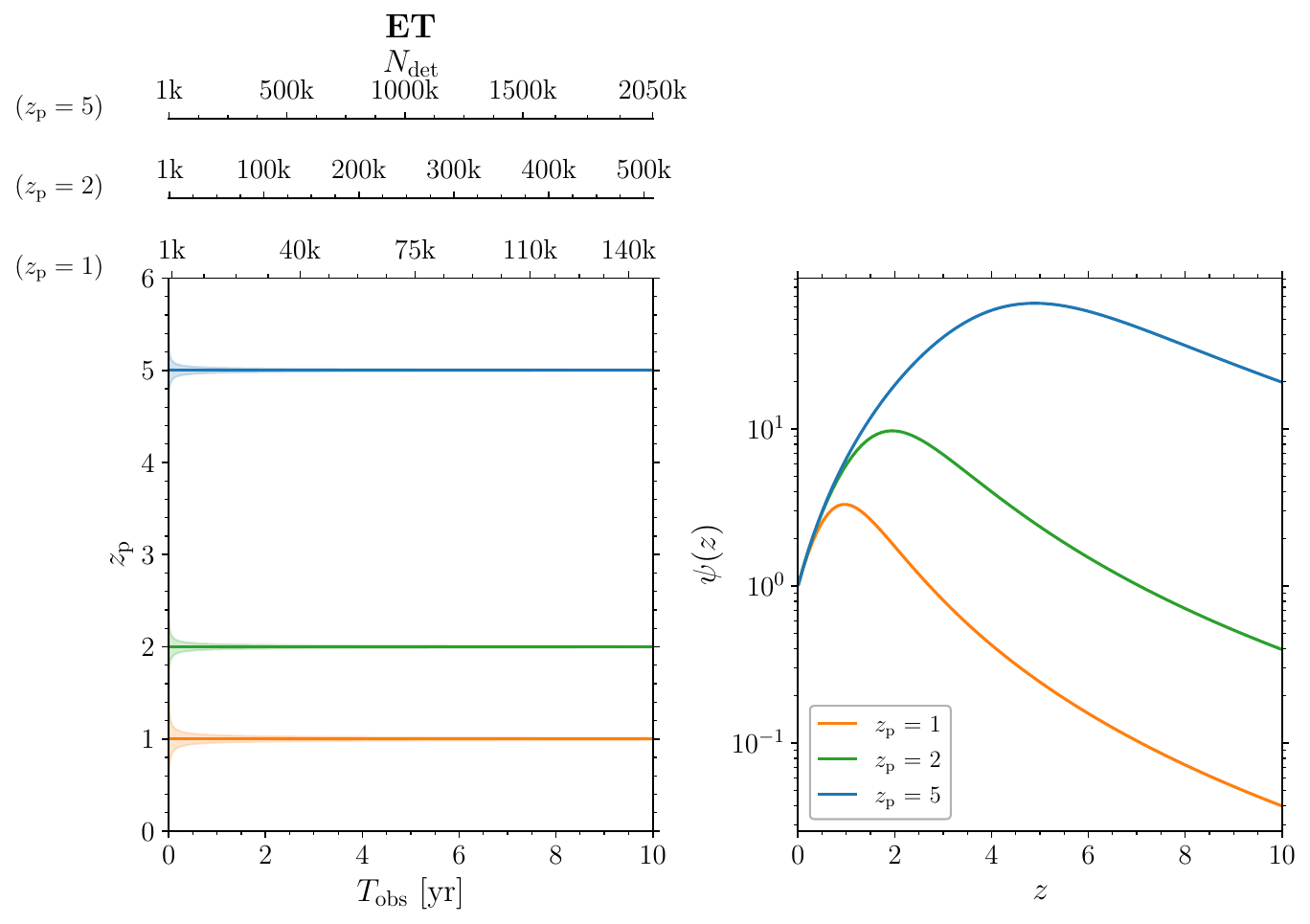}
    \caption{Measurability of the redshift at which the merger rate is maximum  $z_{\rm p}$ as a function of the observation time $T_{\rm obs}$ (bottom $x$-axis) for a single ET instrument. We consider three models with $z_{\rm p}=1,\,2,$ and $5$; the three top $x$-axes show the corresponding number of detected events $N_{\rm det}$. The fiducial model is represented by the green curve with $z_{\rm p}=2$. 
 The left panel show the injected values (horizontal lines) and the related $1\sigma$ errors (shaded areas). For each model, the corresponding merger rate $\psi(z)$ is illustrated in the right panels; this is normalized such that $\psi(z=0)=1$.
Related results are presented in Table~\ref{tab:modelvariations}.
       }
        \label{fig:zp_figure}
$\,$\\
\includegraphics[width=0.75\textwidth]{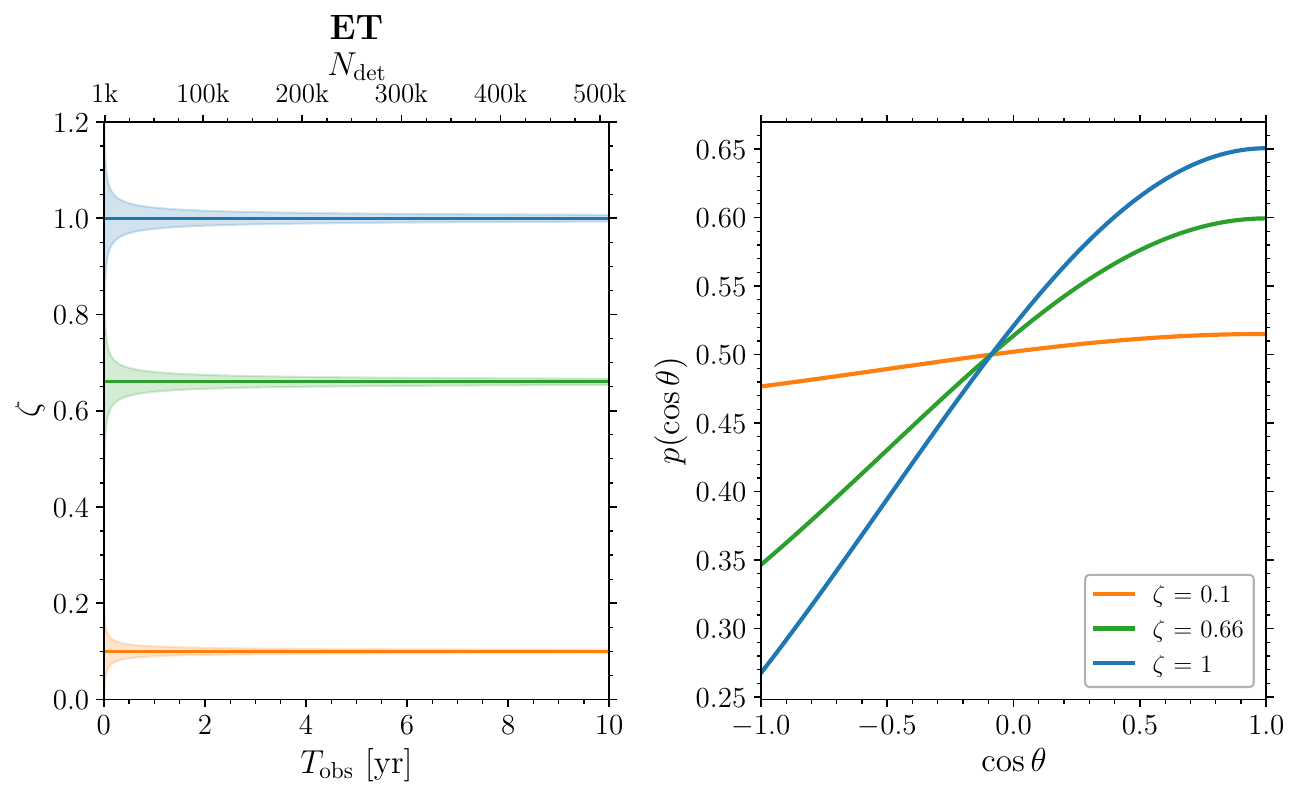}
    \caption{Measurability of the mixing fraction $\zeta$ between BH binaries with aligned and misaligned spins as a function of the observation time $T_{\rm obs}$ (bottom $x$-axis) for a single ET instrument. The corresponding number of detections is reported in the top $x$-axis.
    We consider three model variations with $\zeta=0.1,\,0.66,$ and $1$. The fiducial model is represented by the green curve with $\zeta=0.66$. The left panel shows the injected values (horizontal lines) and the related $1\sigma$ errors (shaded areas). %
    For each model, the corresponding distribution of the spin-orbit angle is illustrated in the right panels.  Related results are presented in Table~\ref{tab:modelvariations}.
        }
        \label{fig:zetachi_figure}
\end{figure*}

There is an important statistical caveat that affects some of the results we just described. Errors on broad features scale as the square root of the number of data-points, which is indeed the prediction of Eq.~\eqref{scaling}: $\sigma_\lambda\propto 1/\sqrt{N_{\rm det}}$. This agrees with one's intuition from the central limit theorem. However, the case of cutoff parameters is different and, instead, errors are expected to scale as $1/{N_{\rm det}}$ (this is the popular ``German tank'' problem in statistics~\cite{2021arXiv210108162C}). This property is not captured by the Fisher matrix formalism, which expands the likelihood around the true value $\bar\lambda$ --- an expansion that should not be allowed at the edge of a distribution. While mitigated by the smoothing parameters $\sigma_{\rm h}$ and $\sigma_{\rm l}$, this caveat affects our Fisher-based error estimates of $m_{\rm min}$ and $m_{\rm max}$. Our results are therefore conservative: we predict a $1/\sqrt{T_{\rm obs}}$ scaling but this could actually be as rapid as $1/{T_{\rm obs}}$. {This same feature was already noted in single-event parameter estimation when experimenting with truncated waveforms~\cite{2014CQGra..31o5005M}.} %

\subsection{Redshift distribution}

Current GW data provide conclusive evidence that the BH merger rate evolves with redshift. Assuming the merger rate evolves as $R(z) = R_0 (1+z)^\kappa$, GWTC-3 data indicate $\kappa \sim 3$~\cite{2023PhRvX..13a1048A}, which is consistent with the rise of the star-formation rate with redshift in the local Universe~\cite{2021ApJ...914L..30F,2022ApJ...931...17V,2019MNRAS.490.3740N,2021MNRAS.502.4877S}. Modeling the evolution of the merger rate with redshift using a single power-law is not appropriate for 3G detectors because their detection horizon extends far beyond the peak of the 
star-formation 
rate. This is indeed the key reason behind choosing the {\sc Madau--Dickinson} profile instead, cf.~Sec.~\ref{popmodels}.

The third row of Fig.~\ref{fig:relative_errors} illustrates the relative errors on the redshift hyperparameters for the fiducial model, assuming either ET or ET+2CE. Again, we report qualitatively similar performances for both detector configurations. However, when compared to masses and spins, the redshift distribution is more prominently affected by the expansion of the detector network. 
This is because a network of well-separated interferometers has better sky localization capabilities, thus breaking the known degeneracy with the distance to the source, and thus with the redshift (let us recall that here we are fixing the cosmology). 
Improvements saturate as $T_{\rm obs}$ increases and marginal gains require longer times. For our fiducial model, we find that all parameters associated with the redshift evolution achieve 1\% accuracy within a range of 2 months to 1.5 years when using ET+2CE, whereas the single ET configuration requires 2 to 4.5 years~(Table~\ref{tab:ET_errors}). 

Figure~\ref{fig:cornerET} shows that all the redshift hyperparameters are tightly correlated. %
In particular, there is a positive (negative) correlation between $\beta_z$ ($\alpha_z$) and the peak of the merger rate $z_{\rm p}$. This is largely because we are keeping the local merger rate $R_0$ fixed. If the redshift peak is higher, the merger rate must be steeper at high redshift and shallower at low redshift.

Finally, Fig.~\ref{fig:zp_figure} shows a model-variation study where we shift the peak of the redshift distribution $z_{\rm p}=1,\,2,\,5$,  considering a single ET instrument. We find that the errors decrease with increasing $z_{\rm p}$, which is once more a direct consequence of our rate choice. Fixing the local rate $R_0$ implies that these different models refer to Universes with a different number of BH mergers $N$ (and thus a different number of detected events $N_{\rm det}$); cf.~App.~\ref{app:redshift}. In this context, assuming a model with a large value of $z_{\rm p}$ corresponds to assuming more merging BHs, which in turn results in better population constraints. An alternative choice would be to fix the total merger rate instead of the local value $R_0$. %

\subsection{Spin distribution}

As for the spin distribution, current GW data reveals evidence of both aligned and misaligned spins. Systems with large spin-orbit misalignment suggest formation in dynamically active environments like globular clusters or AGN disks, where interactions can randomize spin directions; on the other hand, aligned spins suggest formation from isolated binary stars ~\cite{2010CQGra..27k4007M,2016ApJ...832L...2R,2017PhRvD..96b3012T,2018PhRvD..98h4036G}. %
The BH spin magnitudes contain information on the details of stellar collapse including the efficiency of angular momentum transfer between stellar cores and envelopes as well as the occurrence of previous mergers~\cite{2017PhRvD..95l4046G, 2017ApJ...840L..24F,2019ApJ...881L...1F,2020A&A...636A.104B}.

Figure~\ref{fig:relative_errors} and Table~\ref{tab:ET_errors} show that, among the spin hyperparameters, those with tighter constraints are the shape parameters entering the Beta distribution of the spin magnitudes. Both ET and ET+2CE achieve percent-level accuracy for $\alpha_\chi$ and $\beta_\chi$ after just a few months of operation. In contrast, the hyperparameters associated with the spin orientations exhibit much longer timescales to reach similar accuracy levels. In particular, the mixing fraction $\zeta$ requires nearly 10 years of observations for both configurations to achieve an accuracy of 1\%. Note that, for the spin case, the population Fisher terms $\Gamma_{\rm II-V}$ become almost comparable to $\Gamma_{\rm I}$, meaning that the single event errors on the spins are no longer irrelevant for the population analysis.

Our results suggest that the broad program of pinpointing the BH binary formation channel using the spin orientations will face some serious challenges, even in the 3G era.
A crucial element on this point will be the achieved low-frequency sensitivity of ET and CE. For the noise curves used here, we have $f\gtrsim 2$~Hz for ET and $f\gtrsim 5$~Hz for CE. %
Pushing the low-frequency requirement will allow the inspiral phase, where spins are prominent observables, to be better resolved, which impacts both accuracies and event rates. %

The corner plot in Fig.~\ref{fig:cornerET} shows that the hyperparameters modeling the spin magnitudes and those modeling the spin directions are respectively correlated within their own groups. %
In particular, we find a positive correlation in both ($\alpha_\chi- \beta_\chi$) and ($\zeta-\sigma_{t}$). We do not observe any meaningful cross-correlations between hyperparameters modeling the spin magnitudes and those modeling the spin orientations, which is expected being the two distributions independent.

Finally, in Fig.~\ref{fig:zetachi_figure} we study the measurability of 
the relative weight of the isotropic ($\zeta=0$) and aligned ($\zeta=1$) component of the spin-direction population, which is meant to capture the fraction of systems originating from dynamical and isolated channels~\cite{2017PhRvD..96b3012T}. We vary $\zeta=0.1,0.66,1 $ while keeping all other hyperparameters fixed to their fiducial values. %
Table~\ref{tab:modelvariations} shows that varying $\zeta$ does not influence the absolute errors on either $\alpha_\chi$ or $\beta_\chi$ (which agrees with the statement above about these parameters being largely uncorrelated). On the other hand, we observe that the error on $\sigma_t$ raises (lowers) for smaller (larger) values of $\zeta$. This can be traced to the fact that larger values for $\zeta$ predict more events in the Gaussian component, which is then easier to characterize. Finally, we find that the absolute uncertainty on $\zeta$ diminishes for lower values of this hyperparameter. As previously remarked, for the spin case single-event parameter reconstruction is found to be relevant through the terms $\Gamma_{\rm II-V}$. For lower values of $\zeta$ a larger fraction of the events have precessing spins, in which case tilt angles are easier to reconstruct as compared to $\theta_i \sim \pm\pi/2$.  %

\section{Conclusions}
\label{sec:conclusions}

We presented a Fisher-matrix implementation to forecast the population properties of merging BH binaries detectable by 3G observatories such as  ET and CE. The starting point of this paper is the formalism developed by~\citeauthor{2023MNRAS.519.2736G}~\cite{2023MNRAS.519.2736G}. Our results show that future detectors will improve the precision with which we can constrain the population of compact binaries by orders of magnitude compared to LIGO/Virgo, with error bars on the hyperparameters tightening at a rate  $\propto 1/\sqrt{T_{\rm obs}}$.  This is actually a conservative estimate, as some of the edge parameters could actually improve as fast as $1/{T_{\rm obs}}$ (the so-called ``German tank'' problem~\cite{2021arXiv210108162C}).

While prospects are exciting, our analysis also reveals that, at least when modeling the population of masses, spins, and redshift, a single ET instrument of triangular shape and a network of ET and two CE's will return qualitatively similar constraints. We expect similar conclusions to hold true also in the case of two L-shaped ET  detectors \cite{2023JCAP...07..068B}. This point is worth investigating more thoroughly.

We predict that some of the key mass-related hyperparameters, such as the minimum BH mass $m_{\rm min}$ and the spectral index of the primary mass distribution $\alpha_m$ can reach sub-percent accuracy within a few years of observation. The best-measured hyperparameter is $\mu_m$, i.e. the position of the Gaussian peak. This is particularly promising, showing that 3G detectors will be able to deliver precision science on localized features in the population of merging compact objects. %
On the other hand, we predict that the upper edge of the primary mass spectrum $m_{\rm max}$ as well as the fraction of binaries with aligned vs misaligned spins are significantly harder to measure, reaching an accuracy of $1\%$ only after $\gtrsim 8$ years of operations and several hundreds of thousands of events. This questions the feasibility of their related science case, namely pair-instability physics and formation channels, respectively. %

We hope this paper will move the signpost from individual events to populations of events in the context of 3G-detector forecasts. This is particularly timely given the current planning activities regarding the 
 construction of such facilities.
Beside 3G detectors, our tool could also be useful for current LIGO/Virgo analyses where, even though this Fisher approach is unreliable, it could provide some intuition about the response of specific parametric population models. %

Our implementation is built on top of the \textsc{GWFast} code~\raisebox{-1pt}{\href{https://github.com/CosmoStatGW/gwfast}{\includegraphics[width=10pt]{GitHub-Mark.pdf}}}~\footnote{\url{https://github.com/CosmoStatGW/gwfast}}~\cite{2022ApJS..263....2I} %
and will be made publicly available in due course~\raisebox{-1pt}{\href{https://github.com/CosmoStatGW/gwfast/tree/master/gwfast/population}{\includegraphics[width=10pt]{GitHub-Mark.pdf}}}~\footnote{\url{https://github.com/CosmoStatGW/gwfast/tree/master/gwfast/population}}. 
While state-of-the-art for LIGO/Virgo and $\mathcal{O}(100)$ events, the population models implemented here are admittedly too simple for 3G detectors, where the much larger statistics will allow probing far more features %
 in the compact-binary population. 
Third-generation detectors will reveal detailed population structures, which will need to be matched by more flexible  population models.
 Developing forecasts in this direction is left to future work.

\acknowledgments
We thank Andrea Antonelli, Ssohrab Borhanian, Rodrigo Tenorio, Arianna Renzini,  Christopher Moore, and Konstantin Leyde for discussions.
V.D.R., D.G., and C.P. are supported by 
ERC Starting Grant No.~945155--GWmining, 
Cariplo Foundation Grant No.~2021-0555, 
MUR PRIN Grant No.~2022-Z9X4XS, 
MUR Grant ``Progetto Dipartimenti di Eccellenza 2023-2027'' (BiCoQ),
and the ICSC National Research Centre funded by NextGenerationEU.   
V.D.R. is supported by an ``Exchange Extra-EU'' scholarship of the University of Milano-Bicocca.
 F.I. is supported by Swiss National Science Foundation, Grant No.~200020$\_$191957 and the SwissMap National Center for Competence in Research. 
 D.G. is supported by MSCA Fellowships No.~101064542--StochRewind and No.~101149270--ProtoBH.
M.M. and V.D.R. supported by the French government under the France 2030 investment plan, as part of the Initiative d'Excellence d'Aix-Marseille Universit\'e -- A*MIDEX AMX-22-CEI-02.
Computational work was performed at CINECA with allocations 
through INFN and Bicocca.

\appendix

\section{Details of the population models}
\label{app:popmodels}
In this appendix we provide details about the parameterized population models %
used in this paper; a concise list of the relevant hyperparameters is provided in Table~\ref{tab:hyperpartable} together with the adopted fiducial values. We stress that, while here we describe only the adopted models, the developed code can trivially be extended to implement different distributions.
\subsection{Mass population model}%
\label{app:masspopmodel}

\setlength{\tabcolsep}{9pt} %
\begin{table*}[t]
\centering
\begin{tabular}{ccc}
\textbf{Parameter} & \multicolumn{1}{c}{\textbf{Description}} & \textbf{Fiducial }\\
\hline \hline
$\lambda_m$ &
\multicolumn{2}{c}{Mass model: {\sc Power--Law Plus Peak}} \\
\hline
$\alpha_m$ & Spectral index for the power-law of the primary mass distribution. & 3.4 \\
$\mmin$ & Minimum mass of the power-law component of the primary mass distribution. & 9.1~$M_{\odot}$\\
$\mmax$ & Maximum mass of the power-law component of the primary mass distribution. & 87~$M_{\odot}$\\
$\lambda_{\rm peak}$ & Fraction of binary BHs in the Gaussian component. & 0.039\\
$\mu_{m}$ & Mean of the Gaussian component in the primary mass distribution. & 34\\
$\sigma_{m}$ & Width of the Gaussian component in the primary mass distribution. & 3.6\\
$\beta_q$ & Spectral index for the power-law of the secondary mass distribution. & 1.1\\
$\sigma_{\rm l}$ & Width of mass smoothing at the lower end of the mass distribution. & 4.0\\
$\sigma_{\rm h}$ & Width of mass smoothing at the upper end of the mass distribution. & 0.5\\
\hline
$\lambda_z$&
\multicolumn{2}{c}{Redshift model: \textsc{Madau--Dickinson}} \\
\hline
$\alpha_{z}$ & Power-law index governing the rise of the merger rate at low redshift. & 2.7\\
$\beta_{z}$ & Power-law index governing the decline of the merger rate at high redshift. & 3.0 \\
$z_{\rm p}$ & Redshift at which the merger rate peaks. & 2.0 \\
\hline
$\lambda_\chi,\lambda_\theta$ &
\multicolumn{2}{c}{Spin model: \textsc{Default}} \\
\hline
$\alpha_\chi$ & Mean of the Beta distribution of the spin magnitudes. & 1.6\\ %
$\beta_\chi$ & Standard deviation of the Beta distribution of the spin magnitudes. & 4.12\\%0.17
$\zeta$ & Mixing fraction of mergers from the truncated Gaussian component for spin orientations. & 0.66 \\
$\sigma_t$ & Width of the truncated Gaussian component for spin orientations. & 1.5
\end{tabular}
\caption{Summary of the population model parameters and their fiducial values. }
\label{tab:hyperpartable}
\end{table*}

For the BH masses we adopt the \textsc{Power--Law+Peak} model first introduced in~\cite{2021ApJ...913L...7A}. The primary (heavier) object is distributed according to
\begin{align}
    p(&m_1|\lambda_{m}) \propto\, [(1 \!-\! \lambda_{\rm peak})\notag \\
    &\times\mathcal{P}(m_1|-\!\alpha_m, \mmin - \sigma_{\rm l}, \mmax + \sigma_{\rm h}) \notag \\
    & + \lambda_{\rm peak}\mathcal{N}(m_1|\mu_m, \sigma_m)]\, \mathcal{S}(m_1|\mmin-\sigma_{\rm l},\sigma_{\rm l}) \notag \\
    & \times [1 - \mathcal{S}(m_1|\mmax,\sigma_{\rm h})]\,,
\end{align}
where %
$\mathcal{P}(m_1|-\alpha_m, \mmin - \sigma_{\rm l}, \mmax + \sigma_{\rm h})$ is a power-law distribution with spectral index $-\alpha_m$ normalized in the interval $[\mmin - \sigma_{\rm l}, \mmax + \sigma_{\rm h}]$, $\mathcal{N}(m_1|\mu_m, \sigma_m)$ is a normalized Gaussian distribution with mean $\mu_m$ and standard deviation $\sigma_m$. For the smoothing functions $\mathcal{S}$, we design and employ a different prescription compared to that of  Ref.~\cite{2021ApJ...913L...7A,2023PhRvX..13a1048A}. The specific functional form of their smoothing is numerically challenging to treat with automatic differentiation. We also note that their functions only \emph{tend} to zero, without actually reaching the limit value. We use the following polynomial filter, which smoothly evolves from 0 to 1 with vanishing first and second derivatives at the edges:
\begin{align}\label{eq:smoothing}
    &\mathcal{S}(x|x_{0},\sigma_x) =
		\begin{cases}
			 0 & \quad x< x_{0}\,, \\
			 f(x|x_{0},\sigma_x) & \quad x_{0}\leq x \leq x_{0} + \sigma_x \,,\\
			 1 & \quad x > x_{0} + \sigma_x \,,
		\end{cases}
		\\
			&f(x|x_{0},\sigma_x) = \dfrac{(x\!-\!x_{0})^3}{\sigma_x^5} [10 \sigma_x^2 \!-\! 15 \sigma_x (x \!-\! x_{0}) \!+\! 6 (x \!-\! x_{0})^2]\,,
			\label{eq:smoothing2}
	\end{align}
where $x_{0}$ represents the lower edge, below which the filter vanishes, and $\sigma_x$ controls the width of the transition. The above definition of the filter makes our implementation numerically stable when computing derivatives, and further allows to analytically compute the normalization of the chosen mass distribution, %
resulting in a faster numerical evaluation. Within our setting, the filtered mass distribution vanishes at $\mmin-\sigma_{\rm l}$ and at $\mmax+\sigma_{\rm h}$, with the onset of the filtering being at $\mmin$ and $\mmax$.

As for the mass of the secondary object, the distribution is given by a power-law with the modified smoothing function described above
\begin{align}
    p(m_2 |\lambda_{m}, m_1) & \propto \mathcal{P}(m_2|\beta_q, \mmin-\sigma_{\rm l},m_1) \notag \\
    & \times\mathcal{S}(m_2|\mmin-\sigma_{\rm l},\sigma_{\rm l}) \notag \\
    & \times [1 - \mathcal{S}(m_2|\mmax,\sigma_{\rm h})]\,,
\end{align}
which is then normalized (again analytically) in the interval $[\mmin-\sigma_{\rm l},\,m_1]$ depending on $m_1$, ensuring that $\int p(m_1,m_2|\lambda_{m})\dd{m_2} = p(m_1|\lambda_{m})$. %

\subsection{Redshift population model}
\label{app:redshift}
We assume that the merger rate only depends on the redshift $z$, i.e. 
\begin{equation}
\frac{\dd N}{\dd t_s \dd V_c}= R_0\psi(z | \lambda_z)\,,
\label{eq:Rz}
\end{equation}
where 
$N$ is the number of mergers (not the number of detections, see below),
$t_s$ is the source-frame time, $V_c$ is the comoving volume, is $R_0$ the local merger rate at $z=0$, and $\psi(z)$ is a function
such that $\psi(z=0)=1$. For the latter, we adopt the Madau-Dickinson profile~\cite{2014ARA&A..52..415M}, which has the following functional form~\cite{2017ApJ...840...39M}

\begin{equation}\label{eq:madau_dickinson}
    \psi(z|\lambda_z) = \frac{ [(1+z_{\rm p})^{\alpha_z + \beta_z} + 1]\, (1+z)^{\alpha_z}}{(1+z_{\rm p})^{\alpha_z + \beta_z} + (1+z)^{\alpha_z + \beta_z}} \, ,
\end{equation}
where $\alpha_z$ ($\beta_z$) model the rise (decline) of the function at low (high) redshift, and $z_{\rm p}$ models the peak of the merger rate. The redshift distribution of the events is then given by~\cite{2015ApJ...806..263D,2018ApJ...863L..41F}
\begin{equation}
p(z |\lambda_z) \propto
\psi(z|\lambda_z)\dfrac{\dd V_c}{\dd z}\dfrac{1}{1+z}\,.
\label{eq:pz}
\end{equation}

The total number of mergers is given by
\begin{align}
N&=%
\int \!\frac{\dd N}{\dd t_s \dd V_c}\frac{\dd V_c}{\dd z} \frac{\dd t_s}{\dd t_d} \dd z \dd t_d
\notag \\
&= T_{\rm obs} R_0 \int_0^{\infty}\!  \psi(z|\lambda_z) \dfrac{1}{1+z}\dfrac{\dd V_c}{\dd z} \dd z \,,%
\label{eq:mergers2}
\end{align}
where $t_d$ is the time at the detector, $\dd t_s/\dd t_d=1/(1+z)$, and $\int \dd t_d = T_{\rm obs}$ %
is the data taking period. %
In Sec.~\ref{sec:Results}, we use the expressions above to estimate the time $T_{\rm obs}$ needed to observe a given number of detected events $N_{\rm det}$.
In particular, we compute $N$ using Eq.~(\ref{eq:mergers2}) and then approximate $N_{\rm det} \simeq N p_{\rm det}(\lambda)$,  neglecting the expected Poisson fluctuation in the number of observed events.

In this paper, we fix the local merger rate to $R_0=17~\text{Gpc}^{-3}~\text{yr}^{-1}$~\cite{2023PhRvX..13a1048A}; %
 computing the Fisher-matrix error on $R_0$ is a natural extension of this paper and requires generalizing Eq.~(\ref{eq:populationLikelihood}) to the non-marginalized population likelihood (see App.~A of~Ref.~\cite{2023MNRAS.519.2736G}). %

\subsection{Spin population model}
\label{app:spins}

For the spins we again follow~\cite{2021ApJ...913L...7A,2023PhRvX..13a1048A} and adopt their \textsc{Default} model. This features a Beta distribution for the (independent) spin magnitudes
\begin{equation}
    p(\chi_{i}| \lambda_\chi) = {\rm Beta}(\chi_{i}|\alpha_{\chi}, \beta_{\chi})\,,\qquad i=\{1,\,2\}\,,\label{eq:Betadistribution}
\end{equation}
where $\alpha_{\chi}$ and $\beta_{\chi}$ are the shape parameters for the distribution. These can be converted into mean and variance, which are used in some of the GW literature to characterize the spin distribution.

The angles $\theta_{i}$ between each spin and the orbital angular momentum of the binary are drawn (again independently) from 
\begin{equation}
    p(\cos\theta_{i}|\lambda_\theta) = \zeta\,{\cal N}_{\rm tr}(\cos\theta_{i}|1,\sigma_t) + (1-\zeta)\,{\cal I}(\cos\theta_{i})\,,\label{eq:tiltdistribution}
\end{equation}
where ${\cal N}_{\rm tr}(\cos\theta_{i}|1,\sigma_t)$ denotes a truncated Gaussian distribution between $[-1,\,1]$ centered on 1 and with standard deviation $\sigma_t$, and ${\cal I}(\cos\theta_{i})$ denotes an isotropic distribution in $[-1,\,1]$.

\section{LIGO/Virgo vs 3G}
\label{appLIGO}

\begin{figure*}[t]
    \includegraphics[width=0.99\textwidth]{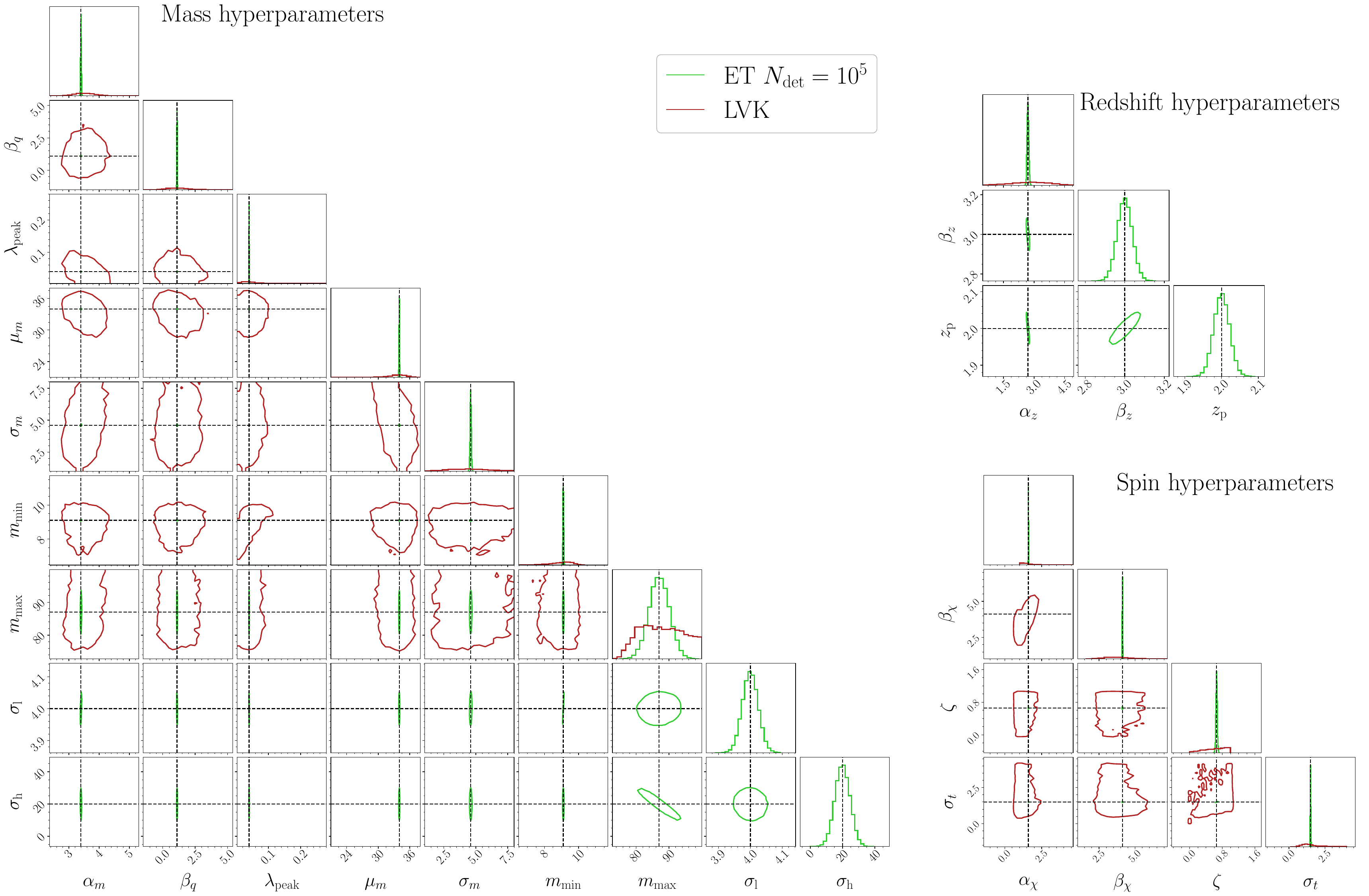}
    \caption{Same as Fig.~\ref{fig:cornerET} but comparing results assuming $N_{\rm det}=10^5$ detections with ET ($T_{\rm obs}\sim 2$~yr, light curves) against current constraints from LIGO/Virgo data (LVK, dark curves). See App.~\ref{appLIGO} for important  caveats.
}   \label{fig:cornerLIGOET}
\end{figure*}

In Fig.~\ref{fig:cornerLIGOET} we compare our results against those obtained with current LIGO/Virgo data. For concreteness, we consider our fiducial model and ET with $N_{\rm det}=10^5$ ($T_{\rm obs}\simeq 2$~yr). We use public LIGO/Virgo/KAGRA data products from GWTC-3~\cite{2023PhRvX..13a1048A}, selecting data from their
{\sc Power--Law Plus Peak} mass distribution, %
{\sc Default} spin model, 
and {\sc Power--Law} redshift model. 

While indicative, there are important caveats to this comparison.
As discussed in App.~\ref{app:masspopmodel},  the smoothing function we adopt differs from that of Ref.~\cite{2023PhRvX..13a1048A}. First, we smooth both the lower and upper ends of the mass spectrum while they only smooth the lower end. The parameters $\sigma_{\rm h}$ and $\sigma_{\rm l}$ have different meanings in the two cases, and are thus omitted from Fig.~\ref{fig:cornerLIGOET}. Moreover, while in our case the mass distribution vanishes at $m \leq m_{\rm min} - \sigma_{\rm l}$, in the prescription used by LIGO/Virgo/KAGRA it instead vanishes at $m \leq m_{\rm min}$. %
For a fairer comparison, we artificially shift the GWTC-3 $m_{\rm min}$ posterior by \( \sigma_{\rm l} = 4~M_\odot \). As for the redshift distribution, the {\sc Power--Law} model used in Ref.~\cite{2023PhRvX..13a1048A} is described by a single hyperparameter  $\kappa$, that can only be approximately compared to $\alpha_z$ from the {\sc Madau--Dickinson}  model.

\bibliography{popfisher}

\end{document}